# Ultra-sensitive integrated circuit sensors based on high-order non-Hermitian topological physics

Wenyuan Deng[1], Wei Zhu[2], Tian Chen[1], Houjun Sun[2], Xiangdong Zhang[1]

[1]*Key Laboratory of advanced optoelectronic quantum architecture and measurements of Ministry of Education, Beijing Key Laboratory of Nanophotonics & Ultrafine Optoelectronic Systems, School of Physics, Beijing Institute of Technology, 100081, Beijing, China*

[2]*Beijing Key Laboratory of Millimeter wave and Terahertz Techniques, School of Information and Electronics, Beijing Institute of Technology, Beijing 100081, China*

**Abstract: High-precision sensors are of fundamental importance in modern society and technology. Although numerous sensors have been developed, obtaining sensors with higher levels of sensitivity and stronger robustness has always been expected. Here, we propose theoretically and demonstrate experimentally a novel class of sensors with superior performances based on exotic properties of high-order non-Hermitian topological physics. The frequency shift induced by perturbations for these sensors can show an exponential growth with respect to the size of the device, which can well beyond the limitations of conventional sensors. The fully integrated circuit chips have been designed and fabricated in a standard 65nm complementary metal oxide semiconductor process technology. The sensitivity of systems not only less than $10^{-3}$fF has been experimentally verified, they are also robust against disorders. Our proposed ultra-sensitive integrated circuit sensors can possess a wide range of applications in various fields and show an exciting prospect for next-generation sensing technologies.**

Sensors with high precision play an important part in many aspects of daily life. There are various schemes for the construction of sensors relying on different physical mechanisms [1-4]. Most sensors rely on resonant structures, where the shifting and splitting of frequency spectra are always used to identify the external perturbation. For example, the photonic microcavity sensor with an ultra-high quality-factor can be used to monitor the change of background refractive index, and the label-free detection of single molecules can be realized [5-7]. The optomechanical transducer can be used as an ultrasensitive detector of weak incoherent forces [8]. In particular, electronic sensors can also offer excellent performances in monitoring multiple environmental parameters [9-15]. Recent advances in the fields of non-Hermitian physics have revealed that enhanced sensitivity can be achieved using a new type of degenerate point, known as exceptional point [16-25]. Furthermore, non-Hermitian topological sensors relying on the anomalous sensitivity to one-dimensional boundary states have also proposed theoretically and demonstrated experimentally [26, 27]. Although numerous sensors have been developed, there is a continuous demand for sensors with increased sensitivity to detect signals that were previously undetectable. In addition, the robust property of sensors has always been a pursuit, which enables them to work in special environments.

On the other hand, higher-order topological phases in Hermitian and non-Hermitian systems have been strongly studied in recent years due to their unconventional physical properties [28-45]. Especially, higher-order non-Hermitian skin effects have also been revealed in some non-Hermitian systems [46, 47]. They lead to new types of boundary physics, which are in contrast to skin modes in the one-order non-Hermitian systems. These higher-order skin effects originate from intrinsic non-Hermitian topology protected by spatial symmetry. The question is whether these higher-order topological physics can be used to realize the sensors with higher levels of sensitivity and stronger robustness.

Here, we propose theoretically a novel class of sensors with superior performances based on exotic properties of high-order non-Hermitian topological physics, the corresponding integrated circuit sensors are fabricated using a 65nm complementary metal oxide semiconductor (CMOS) process technology. Based on non-Hermitian topological corner states, we demonstrate strong weak signal detection characteristics on this platform. Due to the high-frequency oscillation characteristics of the system and advanced nanotechnology, our system can maintain an error of less than 1% at frequencies up to 2GHz. For the first time, we also introduced a Field Programmable Gate Array (FPGA) module to control the non-Hermitian sensing system for achieving high-precision detection of unpredicted measurand. Our work paves the way for ultra-sensitive sensors with stronger robustness.

**Model on high-order non-Hermitian topological sensors**

Now, we provide the model and theory of non-Hermitian high-order topological sensors. We first consider two-dimensional (2D) second-order cases with size of $L_X \times L_Y$, as shown in Fig. 1**a**. The magenta spheres in Fig. 1a represent unit-cells, each unit-cell consists of two lattices (1 and 2) as shown in Fig. 1b. The marks $\lambda_{x(y)}$ and $\lambda'_{x(y)}$ ($\lambda_{x(y)} \neq \lambda'_{x(y)}$) in Fig. 1b display the non-reciprocal couplings among the lattices and unit-cells. By introducing the Fourier transform and Pauli matrices $\sigma_i$ ($i = x, y, z$), the Hamiltonian of the second-order system in $\vec{k}$ space is

$$H = (\lambda_x + \lambda'_x)\cos k_x \sigma_0 + [(\lambda_y + \lambda'_y)\cos k_y + i(\lambda_y - \lambda'_y)\sin k_y]\sigma_x + i(\lambda_x - \lambda'_x)\sin k_x \sigma_z, \quad (1)$$

where $\sigma_0$ is the identity matrix. In open boundary condition (OBC), we can calculate the energy spectrum of the system, as shown in Fig. 1c. It should be noted that the skin mode appears at zero energy in the complex space spectrum [48, 49], indicated by blue in Fig. 1c. The corresponding density of states (DOS) are plotted in Fig. 1d where the skin effect can be clearly observed at zero energy. Here, $\lambda_x = \lambda_y = 2$, $\lambda'_x = \lambda'_y = 10^{-3}$ and $L_x = L_y = 13$ are taken. According to non-Hermitian topological theory [50-52], the skin effect is topologically protected and conforms to the bulk and boundary correspondence (BBC) of non-Hermitian systems.

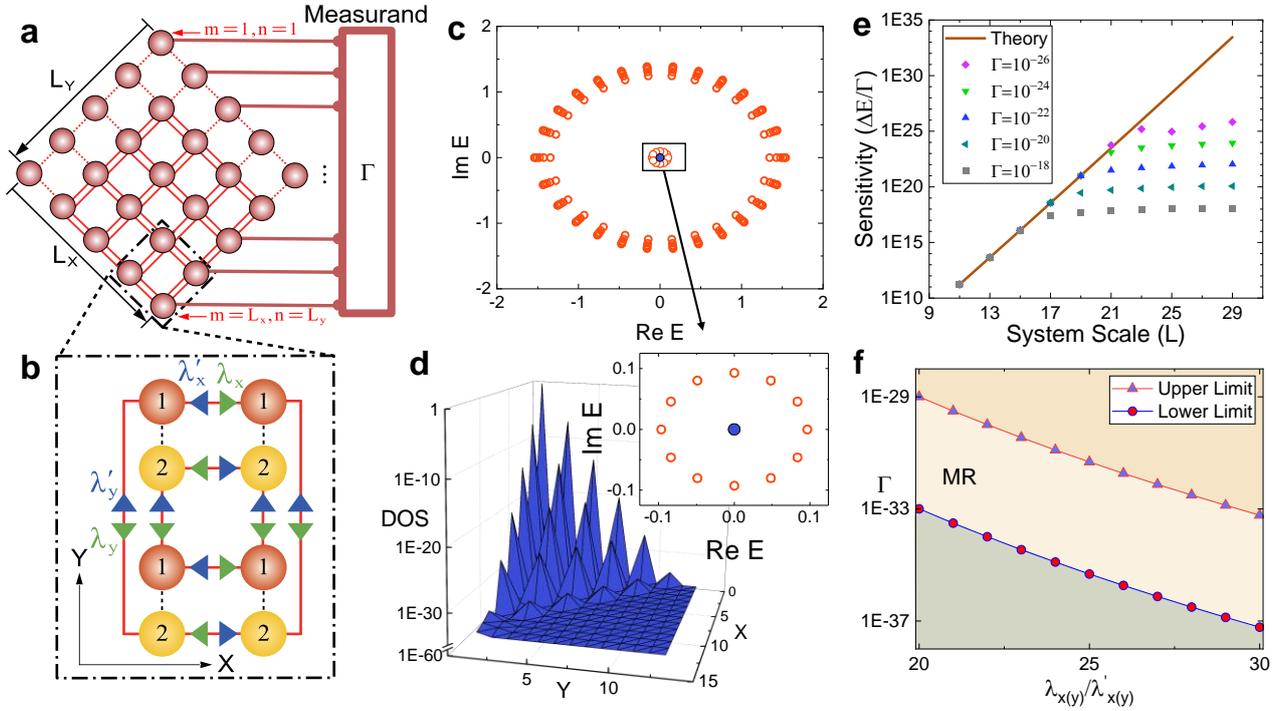

**Figure 1. Second-order non-Hermitian sensing system. a**, Diagram of a second-order sensing model with size of $L_X \times L_Y$. The unit-cells and system couplings are respectively indicated by magenta balls and lines. And the measurand $\Gamma$ indicated by the magenta box can be connected to any two unit-cells. **b**, A part of second-order sensing model including 4 unit-cells indicated by grey box in Fig.1a. The non-Hermitian couplings between cells are indicated by blue ($\lambda_{x(y)}$) and green ($\lambda'_{x(y)}$) arrows. **c**, Energy spectrum of second-order sensing system in complex space where $L_x = L_y = 13$ and $\lambda_x = \lambda_y = 2$, $\lambda'_x = \lambda'_y = 10^{-3}$. The two degenerate zero-energy modes are specially marked with red dots. **d**, Density of states (DOS) of second-order sensing system. The DOS is corresponded to one of the skin modes indicated by blue dots in Fig.

1c. e. Simulation results of the relationship between sensitivity and system scale where $\lambda_x = \lambda_y = 1.9$, $\lambda'_x = \lambda'_y = 0.1$. The brown line is the numerical calculation result of eq. (2) with infinitely small measurand. Other points represent the calculation results of energy spectrum with different measurands $\Gamma = 10^{-18}, 10^{-20}, 10^{-22}, 10^{-24}, 10^{-26}$. **f.** Measurement range for sensing system within measurand $\Gamma$. The triangle and circular dotted lines respectively display the upper and lower limits of system detection, with the middle area representing the measurement range (MR) of the system.

By introducing a measurand $\Gamma$ connected in the above system, as shown in Fig. 1a, the Hamiltonian of the sensing system can be given as $H' = H + H_\Gamma$, where $H_\Gamma = \Gamma(|1,1\rangle\langle m,n| + h.c.)$ represents a weak disturbance and depends on the connection positions of measurand. Here $m$ and $n$ are respectively X and Y coordinates. With $H_\Gamma$ above, the shift of the system skin mode can be obtained by leading order perturbation theory as

$$\Delta E = \frac{\langle \psi_L | H_\Gamma | \psi_R \rangle}{\langle \psi_L | \psi_R \rangle} \approx \frac{\left[2 - \left(\frac{\lambda_x}{\lambda'_x}\right)^{\chi_x} \cdot \left(\frac{\lambda_y}{\lambda'_y}\right)^{\chi_y} - \left(\frac{\lambda_x}{\lambda'_x}\right)^{-\chi_x} \cdot \left(\frac{\lambda_y}{\lambda'_y}\right)^{-\chi_y}\right]}{(\chi_x + 1) \cdot (\chi_y + 1)} \cdot \Gamma$$

$$\xrightarrow{\lambda_{x(y)} > \lambda'_{x(y)} \text{ and } \Gamma \to 0} Ce^K \cdot \Gamma , \qquad (2)$$

where $K = \kappa_x \cdot \chi_x + \kappa_y \cdot \chi_y$, $\kappa_{x(y)} = \ln\left(\frac{\lambda_{x(y)}}{\lambda'_{x(y)}}\right)$, $\chi_{x(y)} = \frac{m(n)-1}{2}$ and $C = \frac{1}{(\chi_x+1)(\chi_y+1)}$. The $|\psi_R\rangle$ and $\langle\psi_L|$ are respectively right and left eigenvectors of skin mode. Detailed derivation for Eq.(2) is given in S1 of Supplementary Material. The sensitivity is generally represented by $\Delta E/\Gamma$ and the brown line in Fig. 1e shows the theoretical results of sensitivity from Eq. (2) as a function of system size $L$. Here the parameters are taken as $m = n = L$, $\lambda_x = \lambda_y = 1.9$ and $\lambda'_x = \lambda'_y = 0.1$. It is an exponential relationship between sensitivity and system scale. And the energy spectrum calculation results are represented by different colored points which are not approximated by $\Gamma \to 0$. When the scale is constant and $\Gamma$ is finite, the sensitivity of second-order non-Hermitian sensing systems has a certain saturation effect with finite measurand. The saturation effect is related to the measurand $\Gamma$, for example, the saturation positions correspond to $L_x = L_y = 17$ and $L_x = L_y = 23$ respectively at $\Gamma = 10^{-18}$ and $\Gamma = 10^{-26}$ as shown in Fig. 1e. The saturation effect and the sensitivity of the system itself determine the upper and lower limits of system detection, respectively. The upper and lower limits of the system are also related to the coupling parameters of the system. Figure 1f displays the relationship between measurand $\Gamma$ and non-Hermitian strength $\left(\frac{\lambda_{x(y)}}{\lambda'_{x(y)}}\right)$ at $L_x = L_y = 13$. As $\frac{\lambda_{x(y)}}{\lambda'_{x(y)}} = 20$, the minimum detectable value is about $10^{-33}$ (lower limit), the corresponding maximum detectable value is about $10^{-29}$

(upper limit). With the increases of the non-Hermitian coupling strength, weaker external measurand can be detected, as indicated by the dots (lower limits) and triangles (upper limits) in Fig. 1**f**. The region between two lines represents the measurement range, which is marked by MR. The MR in Fig. 1**f** is under special parameters, but in fact, we can achieve any MR by adjusting the system scale and coupling parameters. This means that no matter how weak the measurand is, the required sensitivity can be achieved by expanding the scale of the system.

The above only provides second-order results, but in fact, the theory can be extended to third-order and even Nth-order cases by introducing more symmetries. In these higher-order cases, more stronger skin effects can be observed, the sensitivity of the system can be further experimentally improved. The detailed discussions are given in S2 of Supplementary Material. In the following, we explore how to implement the above ultra-sensitive sensors in integrated circuit systems.

# Integrated circuit sensors with ultra-sensitivity based on high-order non-Hermitian topological physics

In order to flexibly change the system size, we prepare a sensor circuit using modular design. Figure 2**a** shows two module units, each corresponds to the theoretical model with a size of 3×3 as shown in Fig.1**a**. The yellow and orange spheres in the module represent the nodes of the circuit network, which correspond to the lattices of the theoretical model. And the nodes are connected through capacitors and buffers (blue and green arrows). Here, the two sensing units are connected and controlled by multiple synchronous switches (SW). In such a case, the theoretical model with a size of 25×25 requires 12 sensing units to be connected in series. In Fig. 2**b**, we show the non-reciprocal couplings among the nodes in the designed circuit system, which correspond to the model shown in Fig.1**b**. The markers 1 and 2 in Fig. 2**b** also correspond to the lattices 1 and 2 in Fig.1**b**. And the non-reciprocal couplings (blue and green arrows) are achieved by controlling the buffer where the capacitor $C_1$ and $C_2$ respectively control the non-reciprocal strength in the corresponding direction. The values of capacitors $C_1$ and $C_2$ correspond to the coupling strengths $\lambda_{x(y)}$ and $\lambda'_{x(y)}$ in Fig. 1**b**. In this way, we can design the circuit network of any size. Additionally, in order to ensure stable operation of the circuit, we have set corresponding grounding capacitors and inductors for each node, and the buffer has also been specially designed accordingly. The grounding method and the principle of buffer are provided in S3 of Supplementary Material. For such a circuit network, we can theoretically derive the circuit Laplacian $J$ and prove that it corresponds to the Hamiltonian $H$ of the lattice model described in Eq. (1). Moreover, the resonance frequency of the oscillator circuit also has the one-to-one correspondence to the eigenvalue of the Hamiltonian $H$. Detailed demonstrations of these relations can be found in S4 of Supplementary Material.

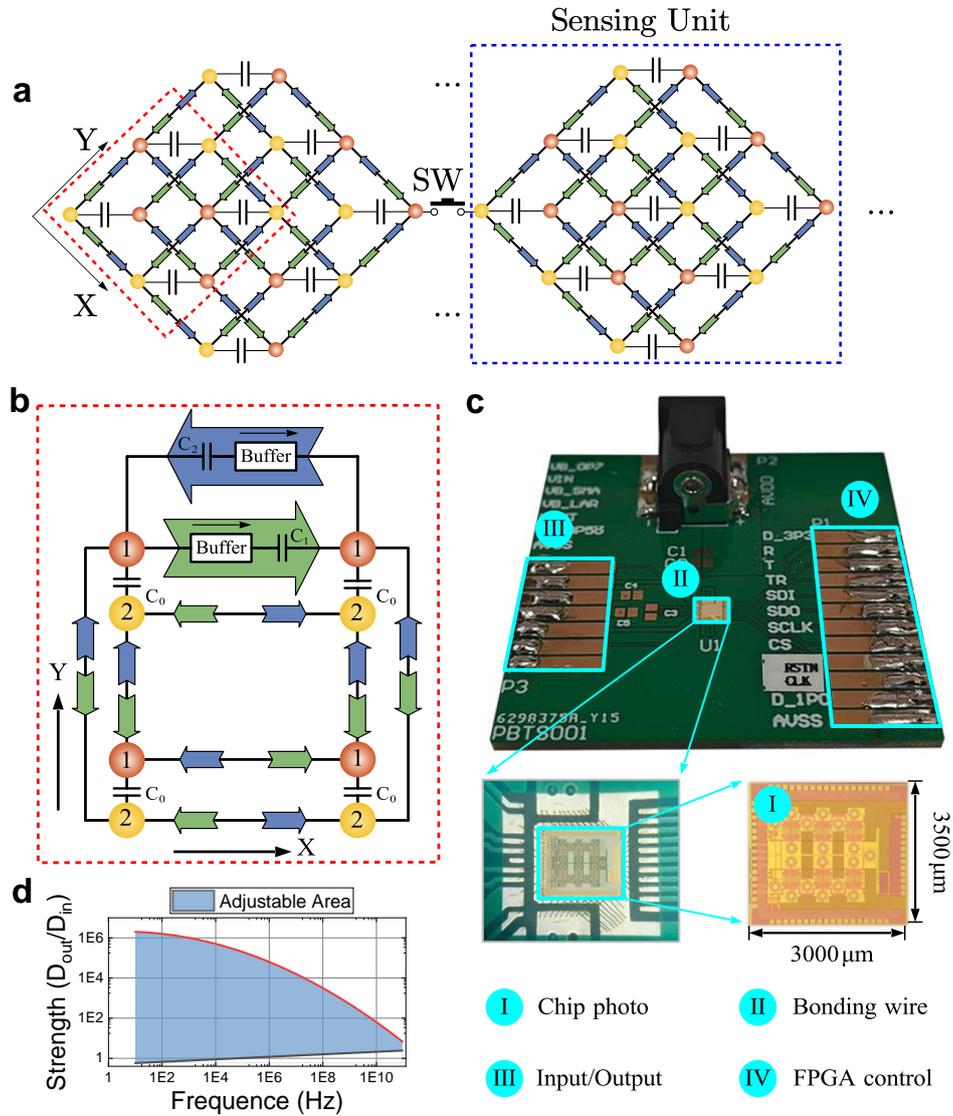

**Figure 2. Illustration and characterization of second-order non-reciprocal sensing chip. a**, The simplified schematic diagram of chip circuit design. The non-reciprocal couplings are indicated by blue and green arrows and $C_0$ represents the capacitor. The red dashed box represents one of the unit-cells in the circuit. The blue dashed box represents the sensing unit of the sensing system, and the sensing units are connected through multiple synchronous switches (SW). **b**, The detailed circuit coupling of sensing system (4 unit-cells). The unit-cell is indicated by magenta which is formed by two lattices connected by capacitor $C_0$. The green ($C_1$) and blue ($C_2$) arrows represent non-reciprocal couplings in different directions, respectively. **c**, Fully integrated non-reciprocal sensing circuit system. The sensing chip is bonded on a printed circuit board (PCB) which provides power supply, input/output interfaces, and FPGA control ports. The chip is fabricated in a 65nm CMOS process technology whose core area is $3000 \times 3500 \mu m^2$. **d**, The simulation results of circuit non-

reciprocal strength range. The non-Hermitian strength ($C_1/C_2$) of the circuit non-Hermitian couplings which is corresponded to the non-Hermitian strength $\lambda_{x(y)}/(\lambda'_{x(y)})$ in quantum models.

Now, we simulate the voltage distribution of the system through Simulation Program with Integrated Circuit Emphasis (SPICE). Detailed simulation method can be found in S5 of Supplementary Material. We consider the case with 12 sensing units, as shown in Fig. 3**a**. This 12-unit system has a total of 13 measurable nodes, marked by Nodes 1 to 13. The resonance frequency for such a LC oscillation circuit is at $f_0 \approx \frac{1}{2\pi\sqrt{2LC_1}}$ which is known as eigenfrequency. If the input signal with $f_0$ is given at Node-13, we can obtain the voltage distribution of the circuit network and the corresponding simulation results indicated by blue line are shown in Fig. 3**b**. Here, $\frac{C_1}{C_2} = 16, C_1 = 5\text{pF}, L = 1\mu\text{H}$ and the corresponding eigenfrequency is $f_0 \approx 1.59\text{GHz}$. It can be seen that the node voltage shows an exponential decreasing trend in circuit system which is completely consistent with the high-order skin effect exhibited in the theoretical model. The orange line represents the simulation results of the node voltage at a randomly selected frequency $f_1$. Comparing the results at frequencies $f_0$ and $f_1$, the highest voltage difference between the two cases reaches 66dB, indicating that non-Hermitian circuit exhibits stronger skin effect at the eigenfrequency.

By utilizing the strong skin effect at the eigenfrequency, we can measure the external measurands through the non-Hermitian sensing circuit. Through the SW controlled by FPGA, we can adjust the position of measurand $C_\Gamma$ in the circuit, as shown in Fig. 3**a**. The nodes $V_{in}$ and $V_{out}$ near the capacitor $C_\Gamma$ are the input and output nodes for the signal, and we can measure the eigenfrequency $f'_0$ of the system with capacitor $C_\Gamma$. And the difference between $f_0$ and $f'_0$ is defined as the eigenfrequency shift $\Delta f = |f_0 - f'_0|$. The relationship between $\Delta f$ and number of working sensing units is provided in Fig. 3c, where the lines of different colors represent the SPICE results for different measurands and $C_1/C_2 = 300$. As the number of sensing units increases, the eigenfrequency shift increases exponentially, increase to a certain extent and show saturation effects, which is completely consistent with the theoretical results in Fig. 1e. Although there are saturation effects, any weaker external measurand can be detected through our designed circuit platform, as long as the non-reciprocal coupling strengths and size of the system are taken appropriately. For example, a frequency shift higher than 1KHz can be obtained for the measurand with $C_\Gamma = 10^{-20} fF$ by using our designed circuit with 12 sensing units. The detailed discussion is given in S6 of Supplementary Material.

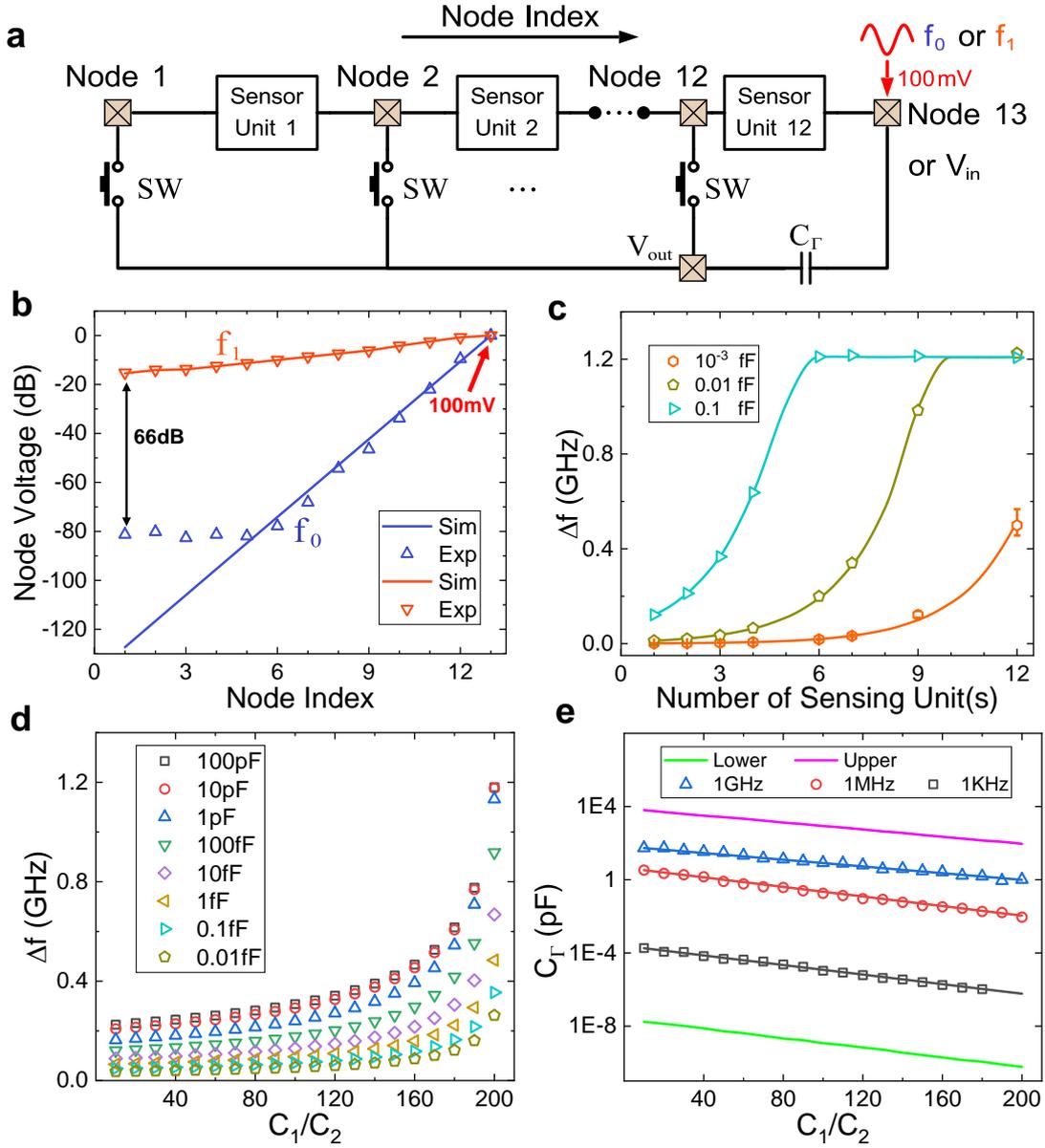

**Figure 3. Skin effect and eigenfrequencies measurement. a,** Schematic diagram of measurement model. The output/input ports are set at the connection positions between the sensing units, marked as node 1 to N. For example, 12 units have 13 nodes. By controlling the switch through FPGA, the access position of external measurand $C_\Gamma$ can be adjusted. Any node can serve as the input and output position for the signal. **b,** Skin effect on the non-reciprocal circuit system with $\frac{C_1}{C_2} = 160, C_1 = 5\text{pF}$ and $L = 1\text{nH}$. The voltage decreases exponentially with the number of units at $f_0 \approx 1.59\text{GHz}$ until a voltage drops of 66dB. The skin effect will decrease at $f_1 \approx 1.27\text{GHz}$. **c.** The relationship between eigenfrequency shift ($\Delta f$) and number of sensing unit(s). Under the control of FPGA, a single chip can achieve non-Hermitian sensing in three states: 1-unit, 3-unit, and 6-unit. By setting the non-reciprocal strength $C_1/C_2 = 300$, the $\Delta f$ increases with the number of sensor units. **d.** The relationship between eigenfrequency shift and non-reciprocal strength ($C_1/C_2$). In the case of

$C_1/C_2 = 10$ to 200 and 6 sensing units, the $\Delta f$ increases exponentially with the increase of $C_1/C_2$. **e**. Measurement range of sensing system under different non-reciprocal strength. Lines and dots represent the theoretical limit measurement boundary and experimental results, respectively.

In the experiment, we integrate the designed circuit system onto the chip by using a standard 65nm CMOS process technology. The inset Ⅰ in Fig. 2**c** shows the microscopic photo of chip which contains six sensing units in an area of 3000×3500 μm². The chip is bonded onto a printed circuit board (PCB) through gold wire (Ⅱ), as shown in Fig. 2**c**. In order to test the function of the chip, we also integrate input/output ports (Ⅲ) on the PCB. In addition, in order to achieve complete controllability of circuit parameters, FPGA ports (Ⅳ) have also been integrated into the PCB as shown in Fig.2**c**. Detailed integration process and design can be found in S7 of Supplementary Material.

The node voltage distributions are measured experimentally at the chip with 12 sensing units under different frequency signals through an oscilloscope spectrum analyzer. Corresponding to the SPICE simulation steps above, we input the signals with $f_0$ and $f_1$ at Node-13, and the measurement results are indicated by dots of different color in Fig. 3b where the error of the experiment does not exceed 1%. Comparing experimental results with the corresponding simulation results, the agreements between them are very well. However, there is also a significant deviation between the experiment and theory when the signal strength is below -80dB. This is due to the influence of environmental thermal noise. When the signal is less than -80dB, the measured signal is suppressed by background noise, and the accurate value cannot be measured. But this phenomenon does not affect the measurement for the eigenfrequency of the system.

Furthermore, we also measure the variation of the eigenfrequency shift with the number of sensing units. As $C_\Gamma = 10^{-3} fF$, the measured results are plotted as orange dots in Fig.3c where $C_1/C_2 = 300$. The pentagonal dots and triangular dots represent the experimental results for $C_\Gamma = 0.01 fF$ and $0.1 fF$, respectively. The good agreements between experimental and simulation results are observed again. However, when the measurand is $C_\Gamma = 10^{-3} fF$, the errors of the experiment slightly exceed the simulation results by 5% to 10%. This is because the tested capacitors produced by two reverse electronic control capacitors has an error of 5% to 10%, which is not a problem with the chips we prepared themselves. If we can have an exact standard capacitor less than $10^{-20} fF$, our prepared chip can also provide good measurement results.

The chip we prepared not only flexibly controls the system size through switches, but also allows us to adjust the non-reciprocal coupling strength of the circuit freely through FPGA. Here, the non-reciprocal coupling strength is roughly defined as $\frac{C_1}{C_2}$ between two nodes, here $\frac{C_1}{C_2} > 1$. Because the non-reciprocal coupling strength is expressed by the capacitors on the chip and is affected by the connected buffer, it has a range of variation, as shown in Fig.2d. In order to control the invariance of other parameters, we take $\frac{C_1}{C_2}$ from 10 to 200 in the experiment, where the system consists of six sensing units and $L = 1\mu H$. Figure 3d shows the experimental measurement results under different values of $\frac{C_1}{C_2}$, and it can be seen that the eigenfrequency shift of the system increases exponentially with the increase of the non-reciprocal coupling strength. And as the measurand increases, saturation effect gradually appears. This can be clearly seen from the experimental results of the system's range as shown in Fig.3e. Here, points and lines represent experimental and simulation results, respectively. And the lower and upper limits represent the simulation results under Δf=1Hz and saturation conditions, respectively. As the non-reciprocal strength changes, the MR of the system exponentially changes which corresponds to theoretical results shown in Fig. 1f. Under this characteristic, the control design of FPGA greatly improves the MR of the non-reciprocal sensing system.

The chip we prepared not only has very high sensitivity, but also has robust characteristics. To verify the robustness of the circuit system, we introduce disordered crosstalk (DC), which is more than 1000 times the actual noise intensity in normal situations. In the experiment, we additionally input DC at the signal input node and define the percentage ratio of DC to the main signal as the strength of DC, such as 10%, 25% and 50%. Figure 4**a** shows the experimental results of voltage skin effect with 50% disordered crosstalk. Here, the parameters are taken identically with those in Fig. 3**b**. When the signal frequency is at the eigenfrequency $f_0$, the skin effect in non-reciprocal system has extremely robust. Even driven by the signal with 50% DC, the skin-effect voltage of the system remains basically unchanged, as indicated by blue (0 DC) and green (50% DC) in Fig. 4**a.** The corresponding experimental working frequencies $f_0$ and $f_1$ are also marked in the figure. Compared to the situation of eigenfrequency, at the frequency of $f_1$, the working signal has been completely distorted, and the skin effect no longer exists. The corresponding sensing experiment results are shown in Fig. 4**b**. It can be seen that under the influence of DC, the sensing performance of the system is not greatly affected and still maintains the characteristic of exponential sensitivity. The inset of Fig. 4**b** shows the corresponding admittance spectrum, indicating that the zero-mode of the system is not affected by DC. The reason that our designed chip has such good robust characteristics is not only due to the high-order topology characteristics, but also because of the band-stop effect of the LC circuit itself. In Fig. 4**c**, we provide experimental measurement results of the band stop effect. Three panels in Fig. 4**c** display the experimental measurement results with $DC =$

−20dB, −7.22dB and − 3.61dB. Strong band-stop effects appear near the eigenfrequency $f_0$ for all cases. This effect helps us reduce noise during the measurement of eigenfrequency, improve system accuracy and robustness. A detailed noise analysis and its mitigation can be found in S8 of Supplementary Material.

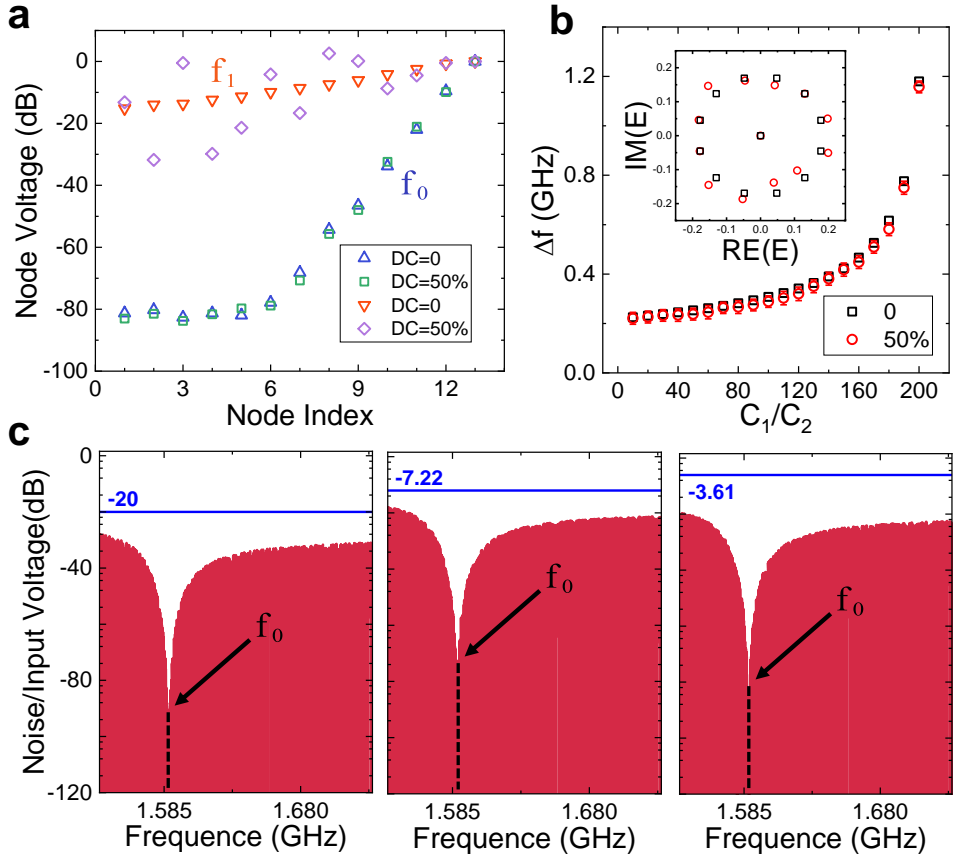

**Figure 4. System disordered crosstalk (DC) test results. a,** The skin effect under DC influence (50%). Here, $C_1/C_2 = 160, C_1 = 5\text{pF}, L = 1\mu\text{H}$. The skin effect remains unchanged at $f_0 \approx 1.59\text{GHz}$ and is not affected by DC interference which is protected by topology. At $f_1 \approx 1.27\text{GHz}$, the system is greatly affected by DC, and the signal tends to become chaotic. **b,** System eigenfrequency shift under DC influence (50%). There will be a deviation of less than 5% in the eigenfrequency shift when DC=50%. The inset represents the energy spectrum solution of the system admittance matrix in this case. **c,** Signal noise measurement results. The circuit parameters are $C_{out}/C_{in} = 160, C_1 = 5\text{pF}, L = 1\text{nH}$ and the theoretically eigenfrequency is 1.59GHz. The red part represents the frequency sweep results of the DC. Here, the crosstalk test is set as DC = −20dB, −7.22dB and − 3.61dB.

## Discussion and conclusion

The designed and fabricated circuit sensors (chips) above only focus on the second-order non-Hermitian case. In fact, the design and fabrication can be extended to third-order, fourth-order, and even more higher-order cases. The sensitivity and robustness of the systems increase exponentially with the order of the system. These high-order non-Hermitian sensors have a strong detection advantage for capacitor front-end which are widely used in our daily lives. At the same time, based on the correspondence between circuits and quantum systems, we can also achieve corresponding detection circuits for resistors and inductors, by adjusting the types of relevant devices used in the circuit. In addition, the strong topology protection characteristics enable the chip to have a high success rate during the preparation process. This is because its higher-order symmetry brings higher performance stability, allowing the system to still have exponential sensitivity improvement even when some parts are damaged.

In conclusion, we have theoretically proposed a novel class of sensors with superior performances based on exotic properties of high-order non-Hermitian topological physics. The corresponding integrated electronic platform has been fabricated using a 65nm CMOS process technology. Based on such a platform, we have demonstrated that the designed sensors not only have extremely high sensitivity but also strong robust properties. This means that they can work in various extreme complex environments. At the same time, the 65nm CMOS process we use makes the system highly integrated and can be controlled by FPGA, greatly improving the practicality of the system.

**Acknowledgements**: This work was supported by the National key R & D Program of China (2022YFA1404900), National Natural Science Foundation of China (No.12234004). **Author contributions:** W. Y. Deng finished the theoretical scheme with the help of T. Chen. W. Zhu finished the design of chip under the supervision of H. J. Sun. W. Y. Deng finished the experimental measurements with the help of W. Zhu. X. D. Zhang initiated and designed this research project. **Competing interests:** The authors declare no competing interests. **Data and materials availability:** All data needed to evaluate the conclusions in the paper are present in the paper and/or the supplementary materials

**Supplementary Materials**

Materials and Methods

Figures S1-S7

**Supplementary Information of**

# Ultra-sensitive integrated circuit sensors based on high-order non-Hermitian topological physics

Wenyuan Deng[1], Wei Zhu[2], Tian Chen[1], Houjun Sun[2], Xiangdong Zhang[1]

[1]*Key Laboratory of advanced optoelectronic quantum architecture and measurements of Ministry of Education, Beijing Key Laboratory of Nanophotonics & Ultrafine Optoelectronic Systems, School of Physics, Beijing Institute of Technology, 100081, Beijing, China*

[2]*Beijing Key Laboratory of Millimeter wave and Terahertz Techniques, School of Information and Electronics, Beijing Institute of Technology, Beijing 100081, China*

## S1. Theory derivation of second-order topological non-Hermitian sensing

In S1.1, we provide relevant theoretical derivations for sensing sensitivity based on second-order skin effect under two-dimensional high-order conditions. And compared to the sensitivity of non-Hermitian sensitivity systems that have been validated in one-dimensional models [26, 27], we find that the sensing forms of the system are more diverse and the sensitivity is higher under two-dimensional conditions. At the same time, in S1.2, we further discuss the topological properties in non-Hermitian sensing, and demonstrate that higher-order topology brings stronger sensing protection properties and provides a wider sensing range.

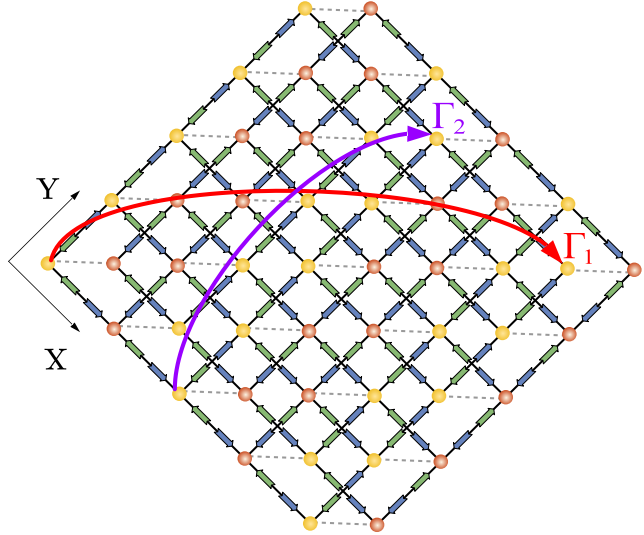

**Supplementary Figure 1: Schematic diagram of a second-order model with size $5 \times 5$.** The system consists of $5 \times 5$ unit-cells, with green and blue arrows respectively indicating non-Hermitian couplings $\lambda'_{x(y)}$ and $\lambda_{x(y)}$. Perturbations $\Gamma_1$ and $\Gamma_2$ are connected to the positions shown in the figure, indicated by purple and red, respectively.

## S1.1 Derivation of sensing sensitivity based on second-order skin effect

In high-order non-Hermitian systems, asymmetric boundary states always occur, for example, the skin effect leads to exponential differences in the distribution of boundary states. In order to construct sharper boundaries, we present a second-order system as shown in Supplementary Figure 1. Here, we provide an example with size of $5 \times 5$, where green and blue arrows represent non-Hermitian couplings $\lambda'_{x(y)}$ and $\lambda_{x(y)}$, respectively. When we do not consider the intracellular coupling represented by dashed lines, the Hamiltonian of the second-order system is expressed as

$$H = (\lambda_x + \lambda'_x)\cos k_x \sigma_0 + [(\lambda_y + \lambda'_y)\cos k_y + i(\lambda_y - \lambda'_y)\sin k_y]\sigma_x + i(\lambda_x - \lambda'_x)\sin k_x \sigma_z, \quad (S3)$$

where $\sigma_0$ is the identity matrix. The skin effect in this system increases exponentially with two orthogonal directions, and the analytical solution shows that one of the zero-mode states of the system is

$$|\psi_R\rangle_{m,n} = \left[\left(\frac{\lambda_y}{\lambda'_y}\right)^{\chi_y} \cdot \left(\frac{\lambda_x}{\lambda'_x}\right)^{\chi_x}, 0, \left(\frac{\lambda_y}{\lambda'_y}\right)^{\chi_y} \cdot \left(\frac{\lambda_x}{\lambda'_x}\right)^{\chi_x-1} \cdots, 1, 0\right]^T_{m,n}, \quad (S4)$$

where $\chi_{x(y)} = \frac{m(n)-1}{2}$, m and n represent the coordinates in the X and Y directions, respectively. For the non-Hermitian system, the left eigenstate is as followed,

$$\langle\psi_L|_{m,n} = \left[\left(\frac{\lambda_y}{\lambda'_y}\right)^{-\chi_y} \cdot \left(\frac{\lambda_x}{\lambda'_x}\right)^{-\chi_x}, 0, \left(\frac{\lambda_y}{\lambda'_y}\right)^{-\chi_y} \cdot \left(\frac{\lambda_x}{\lambda'_x}\right)^{-\chi_x+1}, 0, \cdots, 1, 0\right]_{m,n}. \quad (S5)$$

The zero-mode of the system exhibits an expected exponential difference that the distribution of states shows an exponential trend with size (m and n). In this case, we couple and connect the two non-zero distribution lattices with the greatest difference, such as the lattices whose eigenvalues are $(\lambda_x/\lambda'_x)^{\chi_x} \cdot (\lambda_y/\lambda'_y)^{\chi_y}$ and 1 in Eq. (S2). By introducing a measurand $\Gamma$ connected in the two lattices above, the Hamiltonian of the sensing system can be given as $H' = H + H_\Gamma$, where $H_\Gamma = \Gamma(|1,1\rangle\langle m,n| + h.c.)$ represents a perturbation and depends on the connection positions of the measurand. With $H_\Gamma$ above, the eigenenergy of the system has a certain shift. Through the first-order perturbation approximation, we can obtain the shift of the zero-mode as follows:

$$\Delta E = \frac{\langle\psi_L|H_\Gamma|\psi_R\rangle}{\langle\psi_L|\psi_R\rangle} \approx \frac{\left[\left(\frac{\lambda_x}{\lambda'_x}\right)^{\chi_x} \cdot \left(\frac{\lambda_y}{\lambda'_y}\right)^{\chi_y} + \left(\frac{\lambda_x}{\lambda'_x}\right)^{-\chi_x} \cdot \left(\frac{\lambda_y}{\lambda'_y}\right)^{-\chi_y}\right]}{(\chi_x + 1) \cdot (\chi_y + 1)} \cdot \Gamma$$

$$\xrightarrow{\lambda_{x(y)} > \lambda'_{x(y)} \text{ and } \Gamma \to 0} Ce^K \cdot \Gamma, \quad (S6)$$

where $K = \kappa_x \cdot \chi_x + \kappa_y \cdot \chi_y$, $\kappa_{x(y)} = \ln(\lambda_{x(y)}/\lambda'_{x(y)})$, $\chi_{x(y)} = \frac{m(n)-1}{2}$ and $C = \frac{1}{(\chi_x+1)(\chi_y+1)}$. There is an approximate linear relationship between the zero-mode shift and the measurand $\Gamma$ connected to the system where the linear coefficient has a significant exponential relationship with size and non-Hermitian strength $\lambda_{x(y)}/\lambda'_{x(y)}$. Based on the above derivation, the system has an exponential sensitivity to the external perturbation.

According to the above derivation, the second-order system brings higher sensitivity and provide more methods to connect the perturbation, allowing for perturbation to connect at any lattices on the binary plane. For example, in Supplementary Figure 1, we conduct two methods in a 5 × 5 system, indicated by red and purple arrows, respectively. Utilize Eq. (S4), we can deduce the different energy shifts under two connection methods:

$$\Delta E_1 = \frac{\left[\left(\frac{\lambda_x}{\lambda'_x}\right)^2 \cdot \left(\frac{\lambda_y}{\lambda'_y}\right)^2 + \left(\frac{\lambda_x}{\lambda'_x}\right)^{-2} \cdot \left(\frac{\lambda_y}{\lambda'_y}\right)^{-2}\right]}{9} \cdot \Gamma_1, \tag{S7}$$

$$\Delta E_2 = \frac{\left[\left(\frac{\lambda_x}{\lambda'_x}\right)^1 \cdot \left(\frac{\lambda_y}{\lambda'_y}\right)^1 + \left(\frac{\lambda_x}{\lambda'_x}\right)^{-1} \cdot \left(\frac{\lambda_y}{\lambda'_y}\right)^{-1}\right]}{4} \cdot \Gamma_2. \tag{S8}$$

This property greatly enhances the sensitivity range of the system, which can further enhance the measurement range of the system. Moreover, in the second-order conditions, $\Gamma_1$ and $\Gamma_2$ can be simultaneously connected to the system at different positions to achieve energy shift of $\Delta E = \Delta E_1 + \Delta E_2$. This characteristic brought by high-order systems can be utilized in circuit noise reduction processing, and the detailed methods can be found in S8.

## S1.2 Topological properties in non-Hermitian sensing

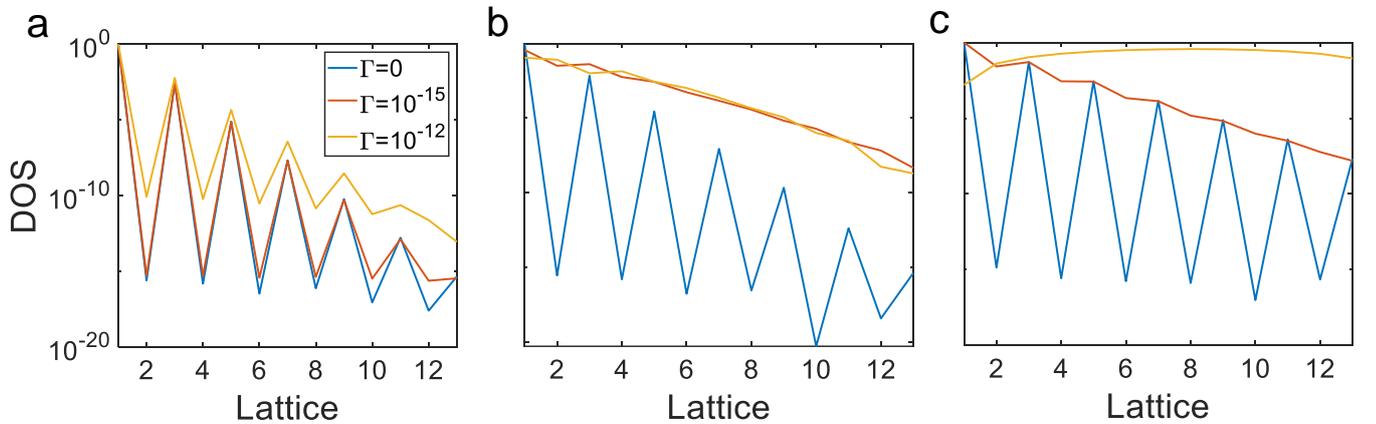

**Supplementary Figure 2: The skin effect of different systems under external perturbations $\Gamma$. a, The skin effect under second-order conditions (zero-mode)**. Different colors correspond to the case where the external perturbation $\Gamma =$

$0, 10^{-15}, 10^{-12}$, respectively. The parameters of the system are $\lambda_x = \lambda_y = 1$, $\lambda'_x = \lambda'_y = 0.9$ and size is $13 \times 13$. **b, The skin effect under second-order conditions (eigenvalue is 0.68).** The parameters are the same to the case of Supplementary Figure 2a. **c, The skin effect under one-dimension conditions (zero-mode).** Different colors correspond to the case where the external perturbation $\Gamma = 0, 10^{-15}, 10^{-12}$, respectively. The parameters of the system are $\lambda_x = 1$, $\lambda'_x = 0.9$ and size is 13.

For any non-Hermitian system, regardless of topology, there are asymmetric state distributions, but not all asymmetric state distributions can be applicable to the sensing method in S1.1. In the derivation of Eq. (S4), we utilize the topological properties of the system and approximately assume that the eigenstates and the topological mode of the system do not change, where $\langle\psi_L|$ and $|\psi_R\rangle$ are both topologically protected. In Supplementary Figure 2a, we present the skin effect of diagonal lattice points in a two-dimensional system under zero-mode. Here, different colors represent the size of perturbations connected to the system, $\Gamma = 0, e^{-12}$ and $e^{-15}$. It can be seen that even with interference of perturbation, the skin effect of the system remains very stable. If the non-Hermitian systems are not topologically protected, such as the skin effect of finite-energy mode or trivial boundary state, the above approximations cannot be used for relevant sensing. In Supplementary Figure 2b, we present the results of the skin effect of diagonal lattice points in a two-dimensional system in the condition of finite-energy mode. The skin effect of the system has seriously shifted when subjected to perturbation interference, and the corresponding system energies are 0.6816 (blue), 0.0504(red), and 0.4936(orange), respectively. Due to the irregular energy shifts, Eq. (S4) cannot be used for the trivial boundary state. Furthermore, Supplementary Figure 2c shows the skin effect with perturbation in one-dimensional topology. Compared to the second-order system, the skin effect of a one-dimensional system exhibits significant deviation with increasing perturbation, and even skin effect disappears completely. Therefore, when designing sensing systems using non-Hermitian skin effect, the topological properties and their strength are the key to improving the stability of the sensing system. A second-order or higher-order system can bring higher system symmetry, thereby making the topological state (sensing performance) of the system more stable.

**S2. Theory for the third-order and nth-order non-Hermitian sensing**

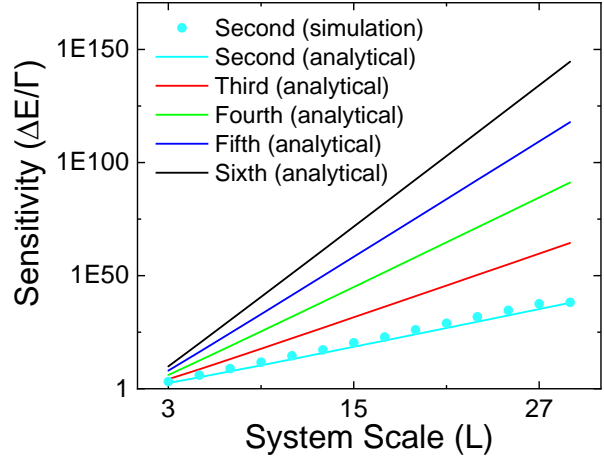

**Supplementary Figure 3: The sensitivity of Nth-order non-Hermitian system.** Different colored lines represent the analytical calculation results of different Nth-order systems. The dots represent the simulation results of the second-order system sensitivity. The parameters are $\lambda_J = 1.9$ and $\lambda'_J = 0.1$.

From the derivation and the approximate conditions in S1, it can be known that the topological properties of the non-Hermitian systems are important in the detection process. Under a unified dimension, the enhancement of the skin effect brought by size is limited. We hope that systems with higher dimensions and orders can break the limitation of system size. For example, we extend the second-order system mentioned above to the three-dimensional cases as follows:

$$H = 2t_1 \cos k_1 \sigma_0 + 2ig_1 \sin k_1 \sigma_z + (2t_2 \cos k_2 + 2ig_2 \sin k_2)\sigma_x + (2t_3 \cos k_3 + 2ig_3 \sin k_3)\sigma_x, \quad (S9)$$

where 1 to 3 represent the three directions. And by the analytic calculation, one of the zero-mode is

$$|\psi_R^3\rangle = \left[\prod_{j=1}^{3}\left(\frac{\lambda_J}{\lambda'_J}\right)^\chi, 0, \cdots, 0, 1, 0\right]^T, \quad (S10)$$

where $\lambda_J = t_J + g_J$ and $\lambda'_J = t_J - g_J$. And its corresponding non-Hermitian left vector is

$$\langle\psi_L^3| = \left[\prod_{j=1}^{3}\left(\frac{\lambda'_J}{\lambda_J}\right)^\chi, \cdots, 1, 0\right]. \quad (S11)$$

According to the second-order theory, in three-dimensional situation, the sensitivity formula of the system to the measurand $\Gamma$ is

$$\Delta E = \frac{\langle\psi_L|H_\Gamma|\psi_R\rangle}{\langle\psi_L|\psi_R\rangle} \approx \frac{\left[2-\prod_{j=1}^{3}\left(\frac{\lambda_J}{\lambda'_J}\right)^\chi - \prod_{j=1}^{3}\left(\frac{\lambda'_J}{\lambda_J}\right)^\chi\right]}{(\chi+1)^3} \cdot \Gamma \xrightarrow{t,g>0;t>g} C_3 \cdot \exp\left(\sum_{j=1}^{3}\kappa_j\right) \cdot \Gamma, \quad (S12)$$

where $\kappa_i = \ln\left(\frac{\lambda_J}{\lambda'_J}\right)$ and $C_3 = \frac{1}{(\chi+1)^3}$. In this situation, we only calculate the maximum sensitivity of the system. It is still obvious that as the order increases, the sensitivity of the system increases exponentially. Now, we can extend the above calculation process to any N-order cases. In the high-order systems designed with special symmetries and topological skin effects, the exponentially-increasing sensitivity of the system remains at

$$\Delta E = \frac{\langle \psi_L | H_\Gamma | \psi_R \rangle}{\langle \psi_L | \psi_R \rangle} \approx \frac{\left[ 2 - \prod_{j=1}^{N} \left( \frac{\lambda_J}{\lambda'_J} \right)^\chi - \prod_{j=1}^{N} \left( \frac{\lambda'_J}{\lambda_J} \right)^\chi \right]}{(\chi+1)^N} \cdot \Gamma \xrightarrow{t,g>0;t>g} C_N \cdot exp\left( \sum_{j=1}^{N} \kappa_j \right) \cdot \Gamma, \quad \text{(S13)}$$

where $C_N = \frac{1}{(\chi+1)^N}$. Supplementary Figure 3 give the analytical results of Eq. (S11), where the $\lambda_J = 1.9$ and $\lambda'_J = 0.1$. In the case of any order, the sensitivity of the system increases exponentially with the size of the system. Comparing the cases of different orders, the sensitivity of the system also increases exponentially with the order of the system. Meanwhile, in order to verify the accuracy of the derivation of Eq. (S11), we have provided simulation results of energy spectrum for the second-order system, as indicated by dots in Supplementary Figure 3. Comparing the analytical and simulation results, the agreements between them are very well. The specific models mentioned above are presented in the article, and this design approach can also be applied to other higher-order models.

## S3. The principle of buffer and the grounding method

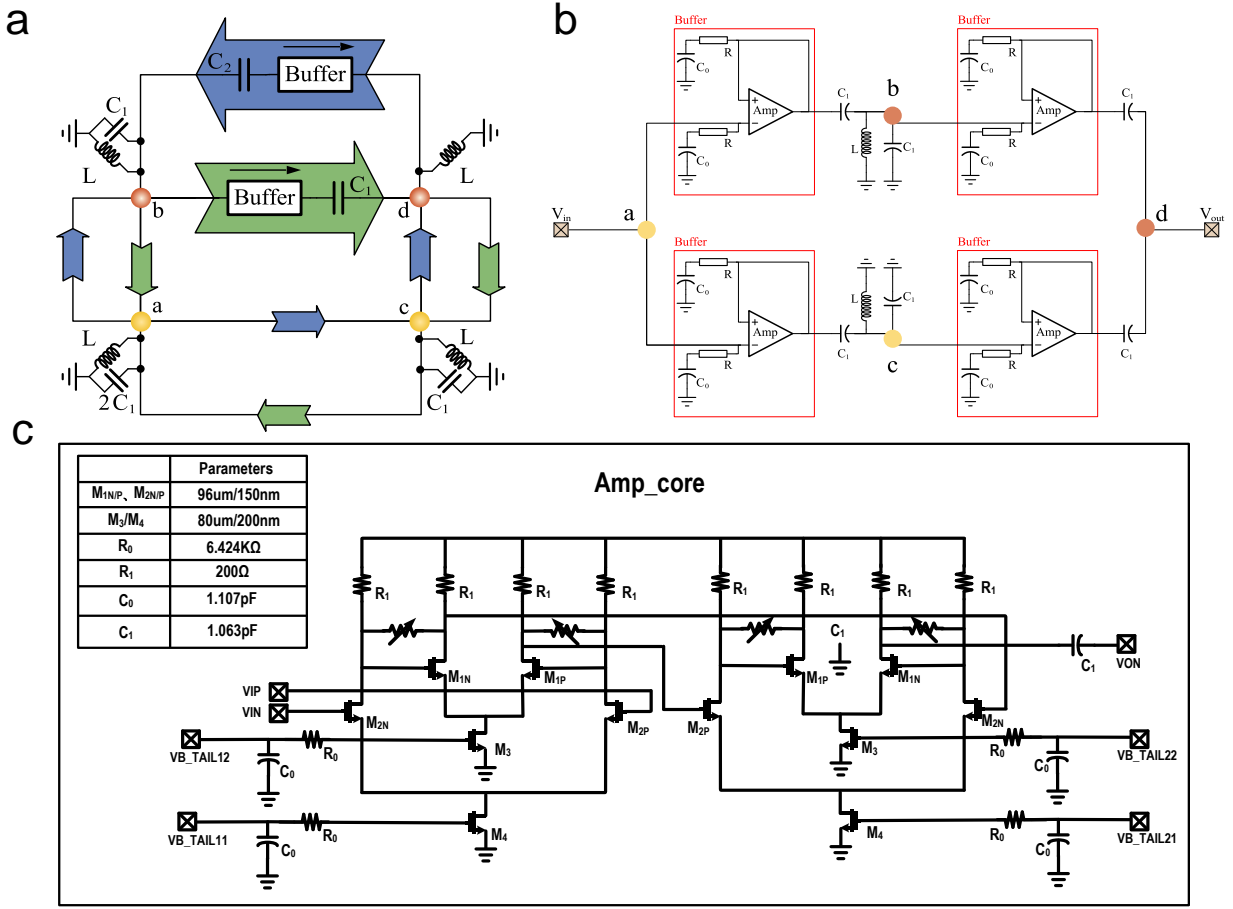

**Supplementary Figure 4: Schematic diagram of second-order sensing circuit grounding and non-Hermitian structure. a**, **The schematic diagram (one sensor unit) of the grounding structure.** The green ($C_1$) and blue ($C_2$) arrows represent non-reciprocal couplings in different directions, respectively. **b**, **The schematic diagram (one sensor unit) of the buffer structure.** The buffer is indicated by a red box, where amplifier (Amp), $C_0$, and R form a high bandwidth and high load voltage follower. **c**, **CMOS design drawings for buffer.**

In sensing circuits, in order to achieve non-reciprocal couplings, active devices are usually needed to introduce non-reciprocal current transmission characteristics, such as commonly used operational amplifiers. However, using an operational amplifier alone can result in a system gain greater than 1, causing noise to be amplified in multiple stages, leading to a decrease in circuit accuracy. To reduce this error, we use a self-designed buffer with non-reciprocal devices, as shown in Supplementary Figure 4a, where $C_1$ and $C_2$ represent the non-reciprocal coupling under the influence of the buffer on the same path, respectively. Here, the forward impedance of the path along the buffer is $C_1$ (green), and the reverse impedance is $C_2$ (blue). Supplementary Figure 4b shows the actual sensing circuit, where $C_1$ is directly adjusted through capacitors, and $C_2$ is regulated through a combination of buffer and capacitor $C_0$. Through the above design, step-by-step control of non-reciprocal circuit coupling $C_1$ and $C_2$ can be achieved on the same path where $C_2$ is defined as the overall reverse imaginary impedance of the buffer. For example, in Supplementary Figure 4b, the admittance from a to b is $i\omega C_1$, while the reverse admittance is $i\omega C_2$, where $C_2$ is much less than $C_1$. Through this method, we can achieve non-reciprocal couplings on the circuit. Moreover, the gain of the buffer is 1, which does not cause multi-level amplification of noise, thereby improving the accuracy of the system. The design drawing of the buffer on the circuit using CMOS technology is shown in Supplementary Figure 4c. By using switches (M) and variable resistors, we can control parameters such as buffer bandwidth, gain, and input/output impedance, which together control the non-reciprocal couplings of the system.

When we use capacitors as circuit couplers, in order to give the circuit resonance characteristics, we need to carry out relevant grounding. The specific grounding methods are shown in Supplementary Figure 4a, with a total of four types. Each method requires the use of the same inductance for grounding, while having different capacitance grounding depending on the coupling situation of the nodes. The above four grounding methods include the grounding forms of all nodes in the entire system, and are selected by using the coupling situation of each node.

## S4. The correspondence between the lattice model and circuit network

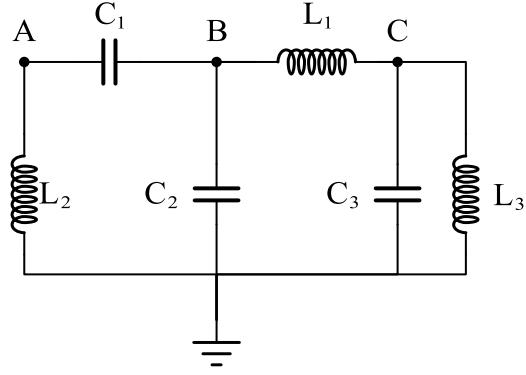

**Supplementary Figure 5: A circuit system consisting of three nodes (A, B, C).** The capacitance and inductance components are represented by C and L respectively. All three nodes A, B and C are grounded by corresponding devices.

### S4.1 Correspondence between circuit component and system coupling

In the circuit, the variation of current and voltage follows Kirchhoff's law. Therefore, for the LC circuit composed of inductors, capacitors and other components, the change behavior of current and voltage is controlled by its circuit Laplace operator, which is similar to that the Hamiltonian describes the energy of physical system.

Taking Supplementary Figure 5 as an example, the circuit network can be represented by a graph whose nodes and edges correspond to circuit nodes and connecting lines or components, respectively. Here, the variation of the node A can be determined by Kirchhoff's current law as

$$I_A = \sum_i C_{Ai}(V_A - V_i) + J_A V_A, \tag{S14}$$

where $I_A$ and $V_A$ are respectively the current flowing into the node A and the potential of the node A and i is the index of the other nodes connected. $C_{Ai}$ represent the capacitors between the node A and other nodes. $J_A$ is the admittance between the node A and the ground. We can use the above theory to construct the topological circuit and show the boundary states in experiments. According to the above theory, the Kirchhoff equations of the three-node system in Supplementary Figure 5 can be given as follows:

$$\begin{cases} I_A = i\omega C_1(V_A - V_B) + \frac{1}{i\omega L_2} V_A \\ I_B = i\omega C_1(V_B - V_A) + \frac{1}{i\omega L_1}(V_B - V_C) + i\omega C_2 \cdot V_B, \\ I_C = \frac{1}{i\omega L_1}(V_C - V_B) + \left(i\omega C_3 + \frac{1}{i\omega L_3}\right) V_C \end{cases} \tag{S15}$$

where $\omega = 2\pi f$ is the circular frequency of input signal. We can write the above equations in matrix form as follows:

$$\begin{bmatrix} I_A \\ I_B \\ I_C \end{bmatrix} = \begin{bmatrix} i\omega C_1 + \frac{1}{i\omega L_2} & -i\omega C_1 & 0 \\ -i\omega C_1 & i\omega(C_1 + C_2) + \frac{1}{i\omega L_1} & -\frac{1}{i\omega L_1} \\ 0 & -\frac{1}{i\omega L_1} & i\omega C_3 + \frac{1}{i\omega L_1} + \frac{1}{i\omega L_3} \end{bmatrix} \begin{bmatrix} V_A \\ V_B \\ V_C \end{bmatrix}. \quad (S16)$$

Furthermore, we can write the above equation in the following form:

$$[\mathbf{A}] = J(\omega) \cdot [\mathbf{V}], \quad (S17)$$

$$J(\omega) = \begin{bmatrix} i\omega C_1 + \frac{1}{i\omega L_2} & -i\omega C_1 & 0 \\ -i\omega C_1 & i\omega(C_1 + C_2) + \frac{1}{i\omega L_1} & -\frac{1}{i\omega L_1} \\ 0 & -\frac{1}{i\omega L_1} & i\omega C_3 + \frac{1}{i\omega L_1} + \frac{1}{i\omega L_3} \end{bmatrix}. \quad (S18)$$

Here, we define $J(\omega)$ as the admittance matrix, which can be used to directly solve the steady-state of the circuit using Laplace transform. And $J(\omega)$ and quantum Hamiltonian have a very similar form, where the lattice points in a quantum system correspond to the nodes in the circuit. And the couplings in a quantum system correspond to the devices between the circuit nodes, such as capacitors and inductors here. Therefore, by adjusting the parameters of each device, we can achieve simulation correspondence for any quantum system.

## S4.2 Correspondence of non-Hermitian system

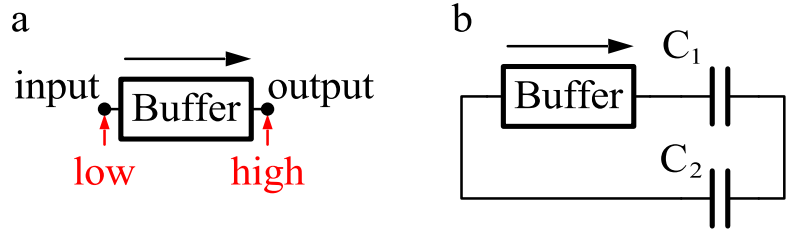

**Supplementary Figure 6: Circuit system of non-Hermitian coupling**. **a, Schematic diagram of buffer input/output impedance**. Arrows indicate the working direction of the buffer where the same direction indicates that the voltage follows the direction, and the output voltage is consistent with the input voltage. The input impedance is close to 0 and the output impedance is close to infinity. **b, Buffer application example**. This circuit achieves non-Hermitian circuit couplings between two nodes.

In the correspondence between circuit admittance matrix and quantum Hamiltonian, capacitors and inductors are both reciprocal devices and can only correspond to Hermitian couplings in quantum models. In S3, we introduced a non-reciprocal circuit device, buffer whose schematic diagram of input and output impedance are shown in Supplementary Figure 6a. The input impedance of the buffer is much lower than the output impedance and tends towards zero impedance, so the forward admittance of the buffer path depends on the admittance of subsequent devices. On the contrary, the output impedance of the buffer is extremely high, causing the reverse admittance of the buffer path to approach zero. This reverse impedance is regulated by the internal components of the buffer and is related to the operating frequency. For example, in Supplementary Figure 6b, we construct two paths, one consisting of buffer and $C_1$, and the other consisting of $C_2$. Based on the properties of the input and output impedance of the buffer and Kirchhoff's current law, we have the following circuit admittance matrix,

$$J(\omega) = -i\omega \begin{bmatrix} -C_1 - C_2 & C_1 + C_2 \\ C_1 & -C_1 \end{bmatrix}. \tag{S19}$$

This admittance matrix achieves non-reciprocal circuit couplings, corresponding to the non-Hermitian couplings in quantum systems. By utilizing the law of non-reciprocal coupling construction, we can add coupling devices on the path with buffer corresponding to non-Hermitian quantum systems. Meanwhile, for Hermitian couplings, we can connect them in parallel to the circuit, like the path with $C_2$.

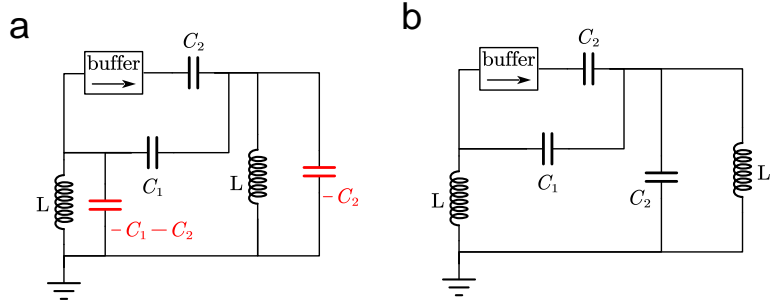

**Supplementary Figure 7: Method of non-Hermitian grounding. a,** Method of negative impedance grounding. **b, Method of redundant capacitors grounding.**

According to Eq. (S17), when constructing the circuit admittance matrix, an unrelated diagonal term $D(\omega)$ is introduced, as follows:

$$J(\omega) = H + i\omega \begin{bmatrix} C_1 + C_2 & 0 \\ 0_1 & C_1 \end{bmatrix} = H + D(\omega). \tag{S20}$$

Here, $H$ is the target Hamiltonian and $D(\omega)$ is the unrelated diagonal term which will cause the system instability and the disappearance of the system's topology. In order to eliminate the influence of diagonal terms, we provide two schemes in the experiment: a. Method of negative impedance grounding; b. Method of redundant capacitors grounding.

In the method of negative impedance grounding, we use capacitors with negative capacitance values for grounding, so that the diagonal terms of the admittance matrix of the circuit system are eliminated, as shown in Supplementary Figure 7a. Following this method, the system admittance matrix is as follows:

$$J_a(\omega) = -i\omega \begin{bmatrix} \frac{1}{\omega^2 L} & C_1 + C_2 \\ C_1 & \frac{1}{\omega^2 L} \end{bmatrix}. \tag{S21}$$

The inductance L is designed to give the circuit resonance characteristics. Through this method, the admittance matrix of the system is completely consistent with the Hamiltonian, and all parameters are corresponded well. However, this method requires the use of negative impedance devices, which are also active devices. The active devices will increase the power consumption of the system. In order to reduce chip power consumption, we adopt the other method using passive devices in the experiment, as shown in Supplementary Figure 7b. The corresponding admittance matrix is as follows:

$$J_b(\omega) = -i\omega \begin{bmatrix} \frac{1}{\omega^2 L} - C_1 - C_2 & C_1 + C_2 \\ C_1 & \frac{1}{\omega^2 L} - C_1 - C_2 \end{bmatrix}. \tag{S22}$$

The method of redundant capacitors grounding makes the diagonal terms of the admittance matrix the same and adds an extra diagonal matrix compared to the Hamiltonian. Here, $D(\omega) = k \cdot I$ where $I$ is the identity matrix and $k = \frac{1}{\omega^2 L} - C_1 - C_2$. The introduction of this identity matrix does not change the topological characteristics of the system spectrum, which does not affect the results of the experiment. By introducing the buffer mentioned above and implementing a special grounding method, we can perform corresponding experiments on any non-Hermitian system in classical circuit.

### S4.3 Correspondence between resonance frequency and eigenvalue

In topological physics, we are concerned about the energy spectrum and topological modes of quantum system. As a result, measuring the eigenenergy of the system is crucial. In classical circuits, the admittance matrix is not directly related to the energy of circuit system, so we need to introduce new measurement methods to observe topological modes in circuit measurements. Based on the discussion in S4.2, we introduce the following general systems for discussing circuit topology modes:

$$J(\omega) = H + D(\omega) = H + (i\omega C - \frac{1}{i\omega L}) \cdot I. \tag{S23}$$

When the quantum Hamiltonian has zero-mode $E_0 = 0$, the determinant of the Hamiltonian is 0 where $|H| = 0$. When the operating frequency of the circuit system is $\omega_0$ and $D(\omega_0) = 0$, the determinant of the circuit admittance matrix is 0. At this time, the operating frequency $\omega_0$ is defined as eigen frequency which is the same to the resonant frequency of the LC circuit. In this case, we can detect the impedance to ground at a certain node in the circuit system, and the theoretical impedance to ground is as follows:

$$[I_g(\omega)] = [J(\omega)]^{-1}, (I_g)_i = \frac{1}{|J(\omega)|} \cdot A_i. \tag{S24}$$

Here, $[I_g(\omega)]$ is the impedance matrix of the system, $(I_g)_i$ is the ground impedance of the node-i and $A_i$ is the algebraic congruent of the impedance corresponding to the node-i. When $J(\omega) = 0$, the ground impedance of each node in the circuit is infinite. Through this principle, we can find the eigenfrequency of the system in experiments by scanning the impedance to the ground, which is completely consistent with the resonant frequency of the LC oscillation system we designed. In the topology circuit constructed using LC oscillation circuit, the zero-mode of the system is related to the resonant frequency (eigenfrequency) of the LC circuit $\omega_0$.

Furthermore, when the zero-mode of the quantum model shifts or the topological mode is finite energy, the above theory can be also applied. We define the energy of the system to be tested as $E$ at this time, $E \neq 0$. In such a case, $|H| \neq 0$, but due to $D(\omega)$ as the identity matrix, we can adjust the operating frequency of the system to $\omega$, making $|H + D(\omega)| = 0$ hold. At this time, $|J(\omega)|$ still equal to 0, so we can still use eq. (S22) to measure the eigenfrequency of the system in the circuit $\omega$. The above measurement methods are applicable to the eigenfrequency of any circuit system.

## S5. Methods of simulation

In the field of integrated electronics, simulating circuits through Simulation Program with Integrated Circuit Emphasis (SPICE) has become very advanced and convenient. For the new non-Hermitian sensing circuit we designed, we have also provided two convenient solutions in the simulation: a. A method of eigenfrequency measurement based on the voltage drop properties caused by skin effect; b. A method of eigenfrequency measurement based on impedance scanning. For different situations, we can use different methods for simulation and experiment. Meanwhile, due to the use of a more advanced 65nm process, the above methods have high practicality in the experiments.

**a. The method of eigenfrequency measurement based on the voltage drop properties caused by skin effect**

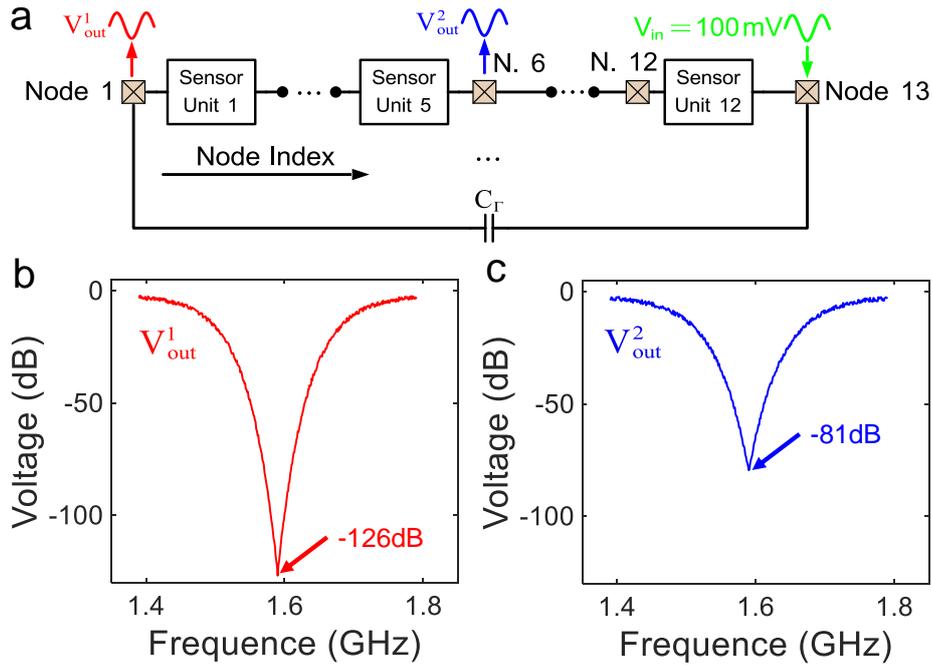

**Supplementary Figure 8: The method of eigenfrequency measurement based on the voltage drop properties caused by skin effect. a, Schematic diagram of measurement model**. The output/input ports are indicated by different colors. And the input voltage is 100mV, using which as a benchmark for measurements. **b, Frequency scan results (simulation) of port $V_{out}^1$.** The simulation frequency scan interval is 1MHz, with a minimum voltage of -126dB where the operating frequency of 1.59GHz. **c, Frequency scan results (simulation) of port $V_{out}^2$.** The simulation frequency scan interval is 1MHz, with a minimum voltage of -81dB where the operating frequency of 1.59GHz.

The non-Hermitian sensing circuit still maintains the skin effect of the non-Hermitian system where the node voltage can exhibit exponential skin effect. By utilizing skin effect, we can perform frequency scanning of voltage at any node of the sensing circuit. The lowest point of voltage in the frequency scanning represents the frequency with the highest intensity of skin effect, which is also the target eigenfrequency.

For example, in Supplementary Figure 8a, we can input a voltage source of 100mV at the node 13 and perform frequency scanning of voltage at nodes 6 and 1, respectively. Here, parameters $C_1/C_2 = 160, C_1 = 5pF$ and $L = 1nH$ are taken. The results of simulated scanning for Node-6 and Node-1 are shown in Supplementary Figure 8b and Supplementary Figure 8c, respectively. The lowest voltage values of scanning results are both at a frequency of 1.59GHz, which is the eigenfrequencies to be measured. Comparing the results of the two cases, as the number of working sensing units increases, the lower the minimum voltage value, the higher the accuracy of the system. In this method, the accuracy of the system depends on the resolution of the signal generator during the frequency scanning and the resolution of the oscilloscope.

**b. A method of eigenfrequency measurement based on impedance scanning**

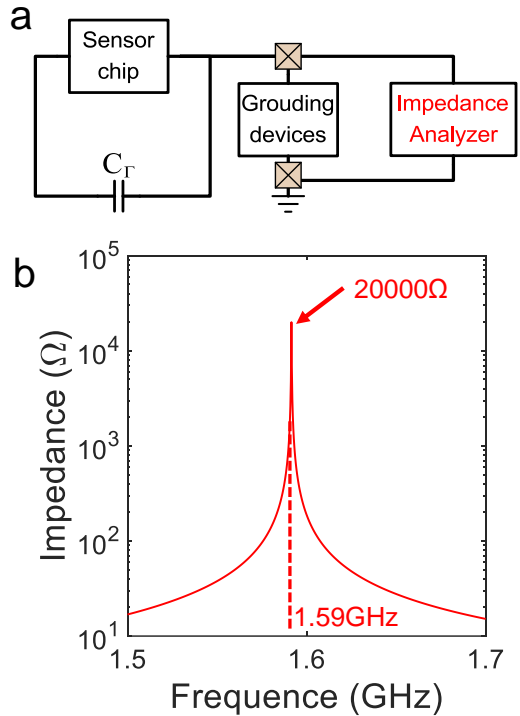

**Supplementary Figure 9: A method of eigenfrequency measurement based on impedance scanning. a, Schematic diagram of measurement model.** The impedance analyzer indicated by red can be connected to any node between ground. **b, Impedance scanning simulation results**. The simulation frequency scan interval is 1MHz, with a maximum impedance of $20000\Omega$ where the operating frequency of 1.59GHz.

According to Eq. (S22), non-Hermitian circuits have extremely high impedance to ground at their eigenfrequencies. By utilizing this feature, we propose the second eigenfrequency measurement method, as shown in Supplementary Figure 9a. In this method, we can use an impedance analyzer indicated by red to perform impedance scanning on any node. Here, parameters $C_1/C_2 = 160, C_1 = 5\text{pF}$ and $L = 1\text{nH}$ still are taken. Supplementary Figure 9b provides the corresponding simulated results, where the peak value of the impedance scanning corresponds to the eigenfrequency of the system. In this method, the accuracy of the system depends on the resolution of the impedance analyzer.

**S6. Extreme detection sensitivity of non-reciprocal sensing chip**

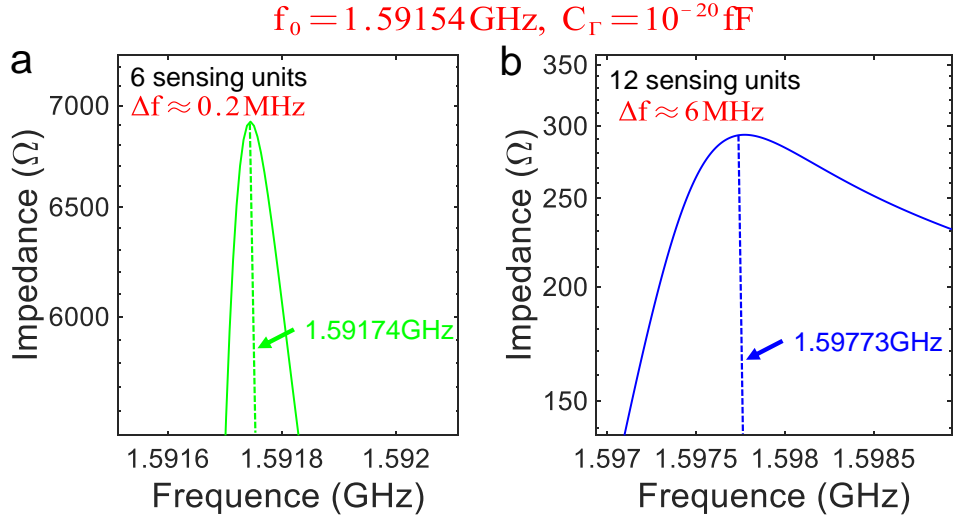

**Supplementary Figure 10: The simulation results of extreme detection sensitivity. a, The simulation results in the case of 6 sensing units.** The measurand $C_\Gamma = 10^{-20}$fF and eigenfrequency of the system $f_0 = 1.59154$GHz. The peak value of impedance scanning is $f = 1.59174$GHz. **b, The simulation results in the case of 12 sensing units.** The measurand $C_\Gamma = 10^{-20}$fF and eigenfrequency of the system $f_0 = 1.59154$GHz. The peak value of impedance scanning is $f = 1.59773$GHz.

In experiments, we provide a measurement result of 0.001fF, and the experimental and simulation data are matched very well. In the measurement experiment of 0.001fF, the frequency shift of the system exceeds 400MHz, which far exceeds the resolution accuracy of the detection instrument. However, due to process limitations, it is not possible to obtain a standard capacitance less than 0.001fF in the experiment, resulting in the inability to calibrate the limit sensitivity. Therefore, we conduct relevant explorations in simulation experiments, where the parameters are taken as $C_1/C_2 = 300, C_1 = 5$pF, $L = 1$nH and $C_\Gamma = 10^{-20}$fF. Supplementary Figures 10a and 10b provide simulation results for 6 and 12 sensing units, respectively. In the case of $C_\Gamma = 10^{-20}$fF, the system still exhibits frequency shift. In the case of 6 sensing units, the frequency offset of the system is about 0.2MHz; in the case of 12 units, the frequency shift of the system is about 6MHz. Therefore, with a 1kHz resolution instrument, this frequency shift can be accurately measured in the experiment.

**S7. Integration process and corresponding methods of measurement**

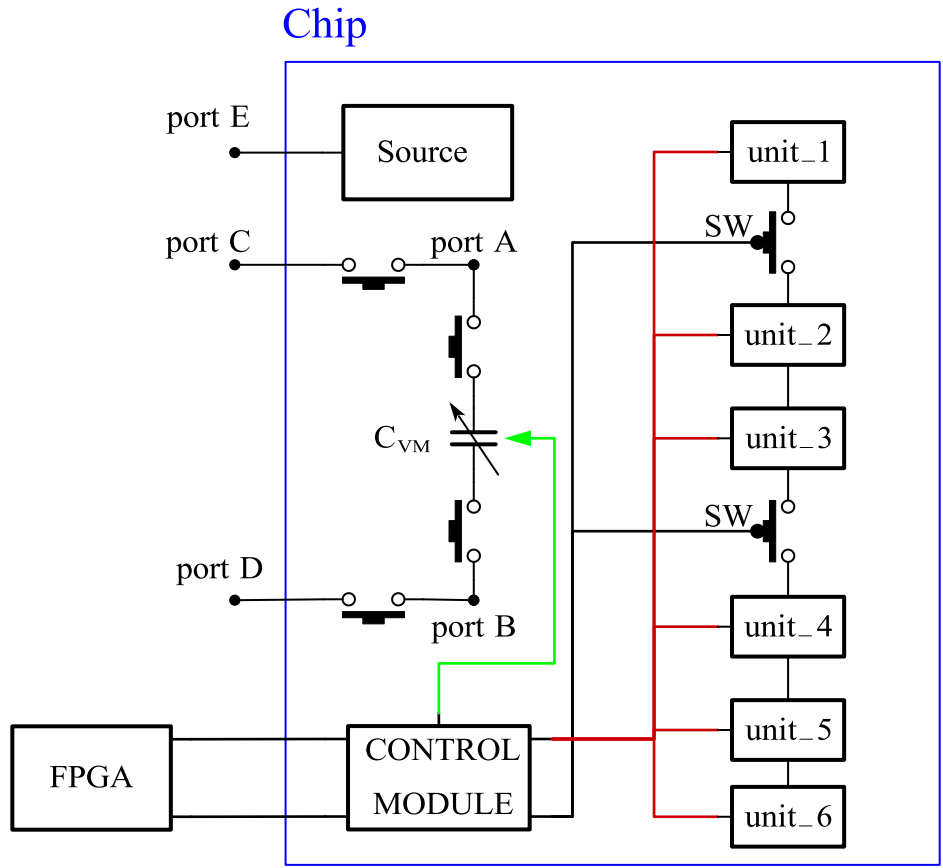

**Supplementary Figure 11: Diagram of chip design and control circuit.** The chip consists of 6 sensing units (indicated by unit_1 to 6), a control module, and a built-in precision capacitor matrix $C_{VM}$. The control module controls the connection mode (SW), sensing coefficient (indicated by red), and precision capacitor parameters (indicated by green). And the chip control module is controlled by an external FPGA. The other ports are the detection interface (C, D) and the power interface (E), respectively.

The sensing system based on the non-Hermitian theory is integrated on one chip with a core area of 3000×3500 $\mu m^2$ using a standard 65nm complementary metal oxide semiconductor (CMOS) process technology. And this chip achieves a $13 \times 13$ non-Hermitian quantum system through the series connection of six sensing units, and the connection method of the system is completely controlled by the external Field Programmable Gate Array (FPGA). As shown in Supplementary Figure 11, for a single chip, there are three connection methods for the sensing unit, namely 1-unit, 3-unit, and 6-unit mode. To achieve the calibration of ultra-sensitive detection, we have embedded an ultra-high precision capacitance matrix ($C_{VM}$) in the chip, which is controlled by FPGA and can achieve a capacitance change of 0.001fF to 60pF (~1%). By connecting the capacitor to the sensing series system, calibration of relevant sensing parameters can be achieved, and we also provide ports (indicated by C, D) that can detect the external capacitance of the chip.

Furthermore, we provide a detailed description of the sensing unit (S7.1) and the method of experiment (S7.2).

## S7.1 Detailed design of sensing unit

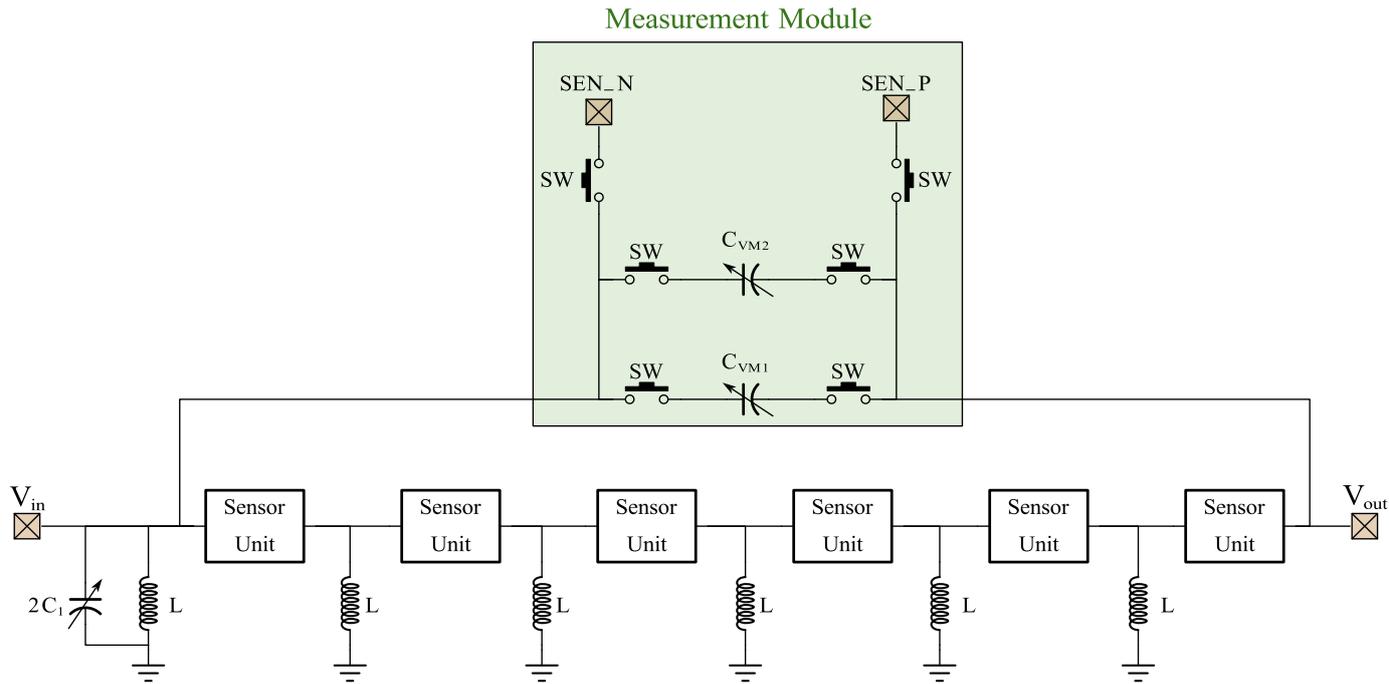

**Supplementary Figure 12: Schematic diagram of non-Hermitian sensing system.** Six sensing units are connected in series, with the number of series units controlled by the FPGA. The ports $V_{in}$ and $V_{out}$ are signal input and output interfaces, respectively. The sensing system is connected to the measurement module directly, where the built-in ultra-high precision capacitance matrix is composed of $C_{VM1}$ and $C_{VM2}$ in parallel, controlled by four switches, achieving no parasitic standard capacitance. And ports SEN_P and SEN_N are respectively external capacitive coupling positive and negative ports which are bonded on PCB.

The overall non-Hermitian sensing system consists of six sensing units connected in series, as shown in Supplementary Figure 12. The voltage signal is input from port $V_{in}$ and processed by the sensing system before being output from the port $V_{out}$, where both ports are bonded to the printed circuit board (PCB). Meanwhile, through these two ports, we can achieve serial connection of two chips (12 units) on the PCB, thereby achieving more sensitive detection systems. We have reserved a grounding capacitor at the input port to reduce parasitic interference on the input signal.

After the sensing system is adjusted using FPGA, we can connect the measured capacitance to the measurement module in the chip, as indicated by green in Supplementary Figure 12. The measurement module is divided into two parts: internal measurement part and external measurement part. The internal measurement is composed of an ultra-high precision capacitance matrix, and the external measurement is achieved by ports bonded on the PCB. In the internal part, two parallel capacitance matrices ($C_{\text{VM1}}$ and $C_{\text{VM2}}$) can provide capacitance ranging from 0.001fF to 100pF (~1%). With this capacitance, we can calibrate the system in the environment where parasitic forces tend to zero. And we have reserved two external ports at the same time, which can be bonded to the PCB. Thus, any capacitance type perturbation can be connected to the chip, but parasitic effects in this state need to be eliminated through other means.

## S7.2 Method of experiment

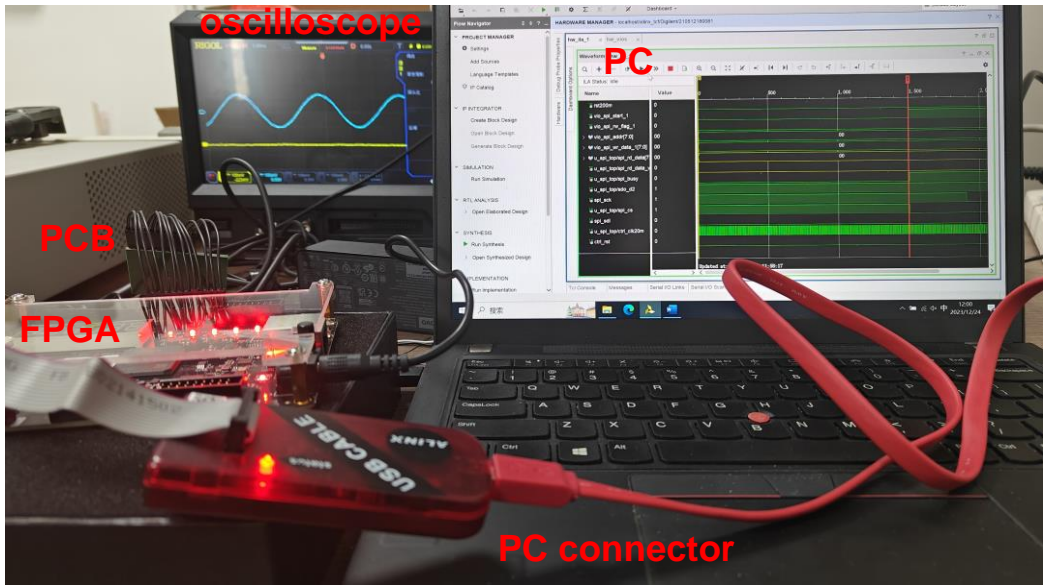

**Supplementary Figure 13: Complete experimental platform**. The platform consists of a PCB, FPGA, and an oscilloscope connected to a personal computer (PC).

In the experiment, we adopt the method of eigenfrequency measurement based on the voltage drop properties caused by skin effect, and the experimental platform is shown in Supplementary Figure 13. After connecting the FPGA and oscilloscope to a personal computer (PC), the experimental device can directly perform related frequency scanning. Compared to the method of impedance scanning, in actual experiments, the power consumption of this method is much lower than the method of impedance scanning. Furthermore, in measurement experiments of greater than 0.001fF, the accuracy of the voltage scanning is sufficient for relevant research. The method based on the voltage drop properties greatly reduces the overall power consumption of the system and reduces the heating problem of the chip. The reduction in power consumption allows us to suppress the thermal noise of the system more easily, enabling the sensing system to operate stably for a long time.

## S8. Analysis on noise sources and mitigation

In the experiment, there are two main sources of noise in the measurement of eigenfrequency shift: a) Multi-level transmission of system thermal noise in active devices; b) The parasitic errors of coupling devices in sensing systems, such as parasitic capacitance, inductance, etc. This section is divided into two parts: suppression of thermal noise (S8.1) and calibration of parasitic errors (S8.2).

## S8.1 Suppression of thermal noise

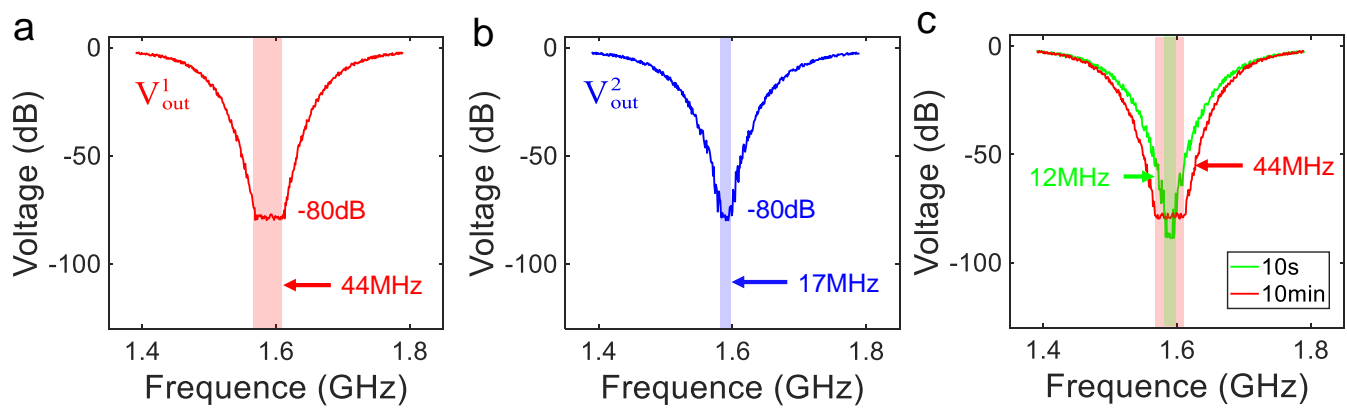

**Supplementary Figure 14: The experimental results of the voltage scanning. a, Frequency scan results of port $V_{out}^1$.** The interval of simulated frequency scanning is 1MHz, with a noise of -80dB where the non-working range is about 44MHz. **b, Frequency scan results of port $V_{out}^2$.** The interval of simulated frequency scanning is 1MHz, with a noise of -80dB where the non-working range is about 17MHz.

In the experiment, environmental thermal noise can cause some background noise in the measurement of the oscilloscope. This background noise can result in a non-working area appearing in the frequency scanning of voltage. The lowest voltage value of the system in this area cannot be accurately obtained, resulting in significant sensing errors in the system. Supplementary Figure 14 shows the experimental results based on the method in Supplementary Figure 8a. Both $V_{out}^1$ and $V_{out}^2$ exhibit certain levels of noise, which leads to non-working regions of 44MHz and 17MHz, respectively. The noise is only related to the temperature of the environment [22], and the working heat of the chip causes a certain amount of increase in the ambient temperature, as shown in Supplementary Figure 14c. Here we present the scanning results of $V_{out}^1$ after ten seconds and ten minutes of operation. The scanning results of ten seconds are from multiple experiments. The results show that as the working time increases, the chip begins to experience heat collection problems, leading to an increase in background noise.

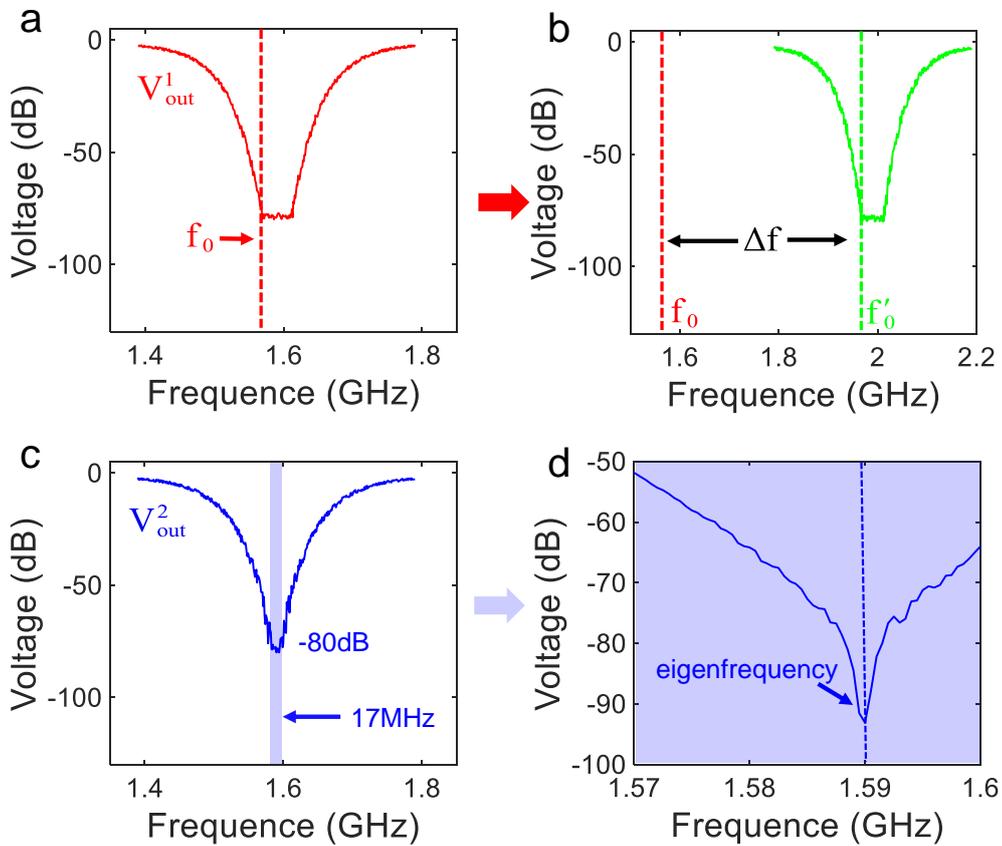

**Supplementary Figure 15: Experimental results with noise. a, Frequency scan results of port $V_{out}^1$.** The interval of simulated frequency scanning is 1MHz, with a noise of -80dB where the non-working range is about 44MHz. **b, Frequency scan results of port $V_{out}^1$ with $C_r = 0.001fF$.** The interval of simulated frequency scanning is 1MHz, with a noise of -80dB where $\Delta f = f_0' - f_0$. **c, Frequency scan results of port $V_{out}^2$.** The interval of simulated frequency scanning is 1MHz, with a noise of -80dB where the non-working range is about 17MHz. **d, Frequency scan results of port $V_{out}^2$ (small range).** The interval of simulated frequency scanning is 1MHz where noise interference decreases.

In order to avoid experimental errors caused by thermal noise, in addition to reducing working time and lowering ambient temperature, we further optimize the experimental methods. For cases where the noise area is too large to completely suppress noise, we conduct the measurement of frequency shift $\Delta f$ using the spline method. Supplementary Figures 15a and 15b provide corresponding experimental results. When it is not possible to accurately determine $f_0$ and $f_0'$, we use the starting frequency of the noise region as $f_0$ and $f_0'$. Under the same environment and working time, the difference of eigenfrequencies after transformation can be approximately estimated as $\Delta f$. Compared with the simulation results, this method can also suppress the error to 1% under noisy conditions. On the other hand, for cases with small noise areas, such as the case of $V_{out}^2$. We can wait for the system to cool down before scanning again in the noisy area. Due to the small scanning area and short working time, the noise is effectively suppressed. Supplementary Figures 15c and 15d provide corresponding experimental results. After further scanning, the system noise is suppressed below the node voltage, resulting in lower measurement errors. The above two methods have been effectively applied in the experiment, and due to the presence of many nodes in the system, we can measure each node and ultimately use the average value as the experimental result.

**S8.2 Calibration of process errors**

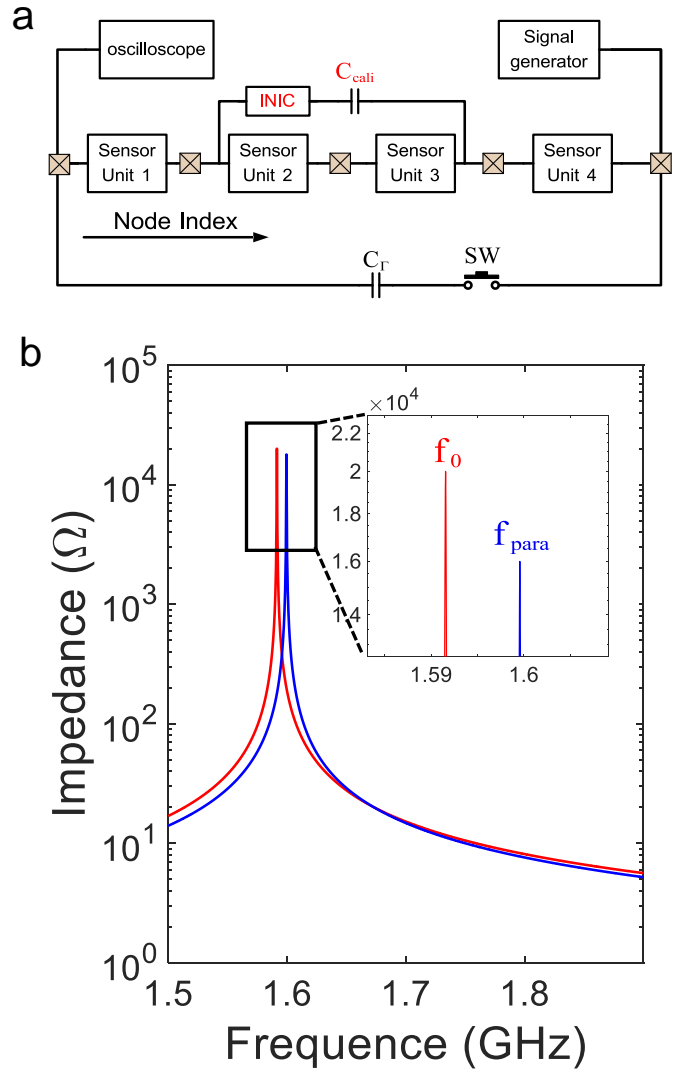

**Supplementary Figure 16: The calibration process of parasitic errors. a, Circuit diagram of the calibration process.** The current negative impedance converter (INIC) and adjustable capacitor ($C_{cali}$) are the calibration unit. **b, The experimental result with parasitic errors.** The red and blue lines represent the disconnection of the detection circuit switch (SW) and the closure of the SW, respectively. Here, the parameter $C_\Gamma = 0$ is still set when the SW is closed.

Supplementary Figure 16a shows a schematic diagram of the four-unit system sensing. When the sensing unit is not connected, that is, when the switch (SW) is turned off, we can measure the eigenfrequency. The measurement results are indicated by the red line in Supplementary Figure 16b, and the frequency corresponding to the impedance peak of the node is $f_0$. However, when we close SW and connect the sensing unit, even though $C_\Gamma = 0$, the system still experiences frequency shift, as indicated by $f_{para}$ in Supplementary Figure 16b. Here, the blue line represents the frequency scanning results ($C_\Gamma = 0$) after the switch is closed. This is because, after the SW is closed, there are parasitic capacitance and inductance in the wires, switches, and other positions of the sensing unit that cannot be ignored under high-frequency conditions, which can cause parasitic error.

To eliminate this error, we introduce a calibration unit in the sensing system, consisting of a current negative impedance converter (INIC) and an adjustable capacitor ($C_{cali}$) highlighted in red in Supplementary Figure 16a. By utilizing the principles of Eq. (S5) and Eq. (S6) in S1.1, we additionally introduce a structure with admittance of $-C_{cali}$. Only when the parameter $-C_{cali}$ is set as $C_{para} - C_{cali} = 0$, the eigenfrequency $f_{para}$ returns to the normal value $f_0$. Therefore, before using the sensing system, we need to perform one calibration to eliminate parasitic errors.